\documentclass[twocolumn]{aastex63}

\usepackage{amsmath}
\usepackage{physics}

\shorttitle{Mapping the Spatially Inhomogeneous Cosmic Reionization}
\shortauthors{Yoshioka et al.}

\begin{document}
\title{CHORUS IV: Mapping the Spatially Inhomogeneous Cosmic Reionization with Subaru HSC}

\author[0000-0002-3800-0554]{Takehiro Yoshioka}
\affiliation{Department of Astronomy, School of Science, The University of Tokyo, 7-3-1 Hongo, Bunkyo-ku, Tokyo, 113-0033, Japan}

\author[0000-0003-3954-4219]{Nobunari Kashikawa}
\affiliation{Department of Astronomy, School of Science, The University of Tokyo, 7-3-1 Hongo, Bunkyo-ku, Tokyo, 113-0033, Japan}
\affiliation{Research Center for the Early Universe, The University of Tokyo, 7-3-1 Hongo, Bunkyo-ku, Tokyo 113-0033, Japan}

\author{Akio K. Inoue}
\affiliation{Department of Physics, School of Advanced Science and Engineering, Faculty of Science and Engineering, Waseda University, 3-4-1 Okubo, Shinjuku, Tokyo 169-8555, Japan}
\affiliation{Waseda Research Institute for Science and Engineering, Faculty of Science and Engineering, Waseda University, 3-4-1 Okubo, Shinjuku, Tokyo 169-8555, Japan}

\author{Satoshi Yamanaka}
\affiliation{General Education Department,
National Institute of Technology, Toba College,
1-1, Ikegami-cho, Toba, Mie 517-8501, Japan}

\author[0000-0002-2597-2231]{Kazuhiro Shimasaku}
\affiliation{Department of Astronomy, School of Science, The University of Tokyo, 7-3-1 Hongo, Bunkyo-ku, Tokyo, 113-0033, Japan}
\affiliation{Research Center for the Early Universe, The University of Tokyo, 7-3-1 Hongo, Bunkyo-ku, Tokyo 113-0033, Japan}

\author[0000-0002-6047-430X]{Yuichi Harikane}
\affiliation{Institute for Cosmic Ray Research, The University of Tokyo, 5-1-5 Kashiwanoha, Kashiwa, Chiba 277-8582, Japan}
\affiliation{Department of Physics and Astronomy, University College London, Gower Street, London WC1E 6BT, UK}

\author{Takatoshi Shibuya}
\affiliation{Kitami Institute of Technology, 165 Koen-cho, Kitami, Hokkaido 090-8507,
Japan}

\author[0000-0002-8857-2905]{Rieko Momose}
\affiliation{Kavli IPMU (WPI), UTIAS, The University of Tokyo, Kashiwa, Chiba 277-8583, Japan}

\author[0000-0002-9453-0381]{Kei Ito}
\affiliation{Department of Astronomical Science, The Graduate University for Advanced Studies, SOKENDAI, Mitaka, Tokyo, 181-8588, Japan}
\affiliation{National Astronomical Observatory of Japan, Mitaka, Tokyo, 181-8588, Japan}
\affiliation{Department of Astronomy, School of Science, The University of Tokyo, 7-3-1 Hongo, Bunkyo-ku, Tokyo, 113-0033, Japan}

\author[0000-0002-2725-302X]{Yongming Liang}
\affiliation{Department of Astronomical Science, The Graduate University for Advanced Studies, SOKENDAI, Mitaka, Tokyo, 181-8588, Japan}
\affiliation{National Astronomical Observatory of Japan, Mitaka, Tokyo, 181-8588, Japan}
\affiliation{Department of Astronomy, School of Science, The University of Tokyo, 7-3-1 Hongo, Bunkyo-ku, Tokyo, 113-0033, Japan}

\author{Rikako Ishimoto}
\affiliation{Department of Astronomy, School of Science, The University of Tokyo, 7-3-1 Hongo, Bunkyo-ku, Tokyo, 113-0033, Japan}

\author{Yoshihiro Takeda}
\affiliation{Department of Astronomy, School of Science, The University of Tokyo, 7-3-1 Hongo, Bunkyo-ku, Tokyo, 113-0033, Japan}

\author[0000-0002-1049-6658]{Masami Ouchi}
\affiliation{National Astronomical Observatory of Japan, Osawa 2-21-1, Mitaka, Tokyo 181-8588, Japan}
\affiliation{Institute for Cosmic Ray Research, The University of Tokyo, Kashiwanoha 5-1-5, Kashiwa, Chiba 277-8582, Japan}
\affiliation{Kavli Institute for the Physics and Mathematics of the Universe (Kavli IPMU, WPI), The University of Tokyo,
Kashiwanoha 5-1-5, Kashiwa, Chiba 277-8583, Japan}

\author[0000-0003-1700-5740]{Chien-Hsiu Lee}
\affiliation{NSF's National Optical-Infrared Astronomy Research Laboratory, 950 N Cherry Ave., Tucson, AZ 86719, USA}

\correspondingauthor{Takehiro Yoshioka}
\email{yoshioka@astron.s.u-tokyo.ac.jp}

\begin{abstract}
The spatial inhomogeneity is one of the important features for understanding the reionization process; however, it has not yet been fully quantified.
To map this inhomogeneous distribution, we simultaneously detect Ly$\alpha$ emitters (LAEs) and Lyman break galaxies (LBGs) at $z \sim 6.6$ from the Subaru/Hyper Suprime-Cam (HSC) large-area ($\sim1.5\,\mathrm{ deg}^2 = 34000\,\mathrm{cMpc}^2$) deep survey.
We estimate the neutral fraction, $x_\mathrm{HI}$, from the observed number density ratio of LAEs to LBGs, $n(\mathrm{LAE})/n(\mathrm{LBG})$ based on numerical radiative transfer {simulation, in which model galaxies are selected to satisfy the observed selection function}.
While the average $x_\mathrm{HI}$ within the field of view is found to be $x_\mathrm{HI} < 0.4$, which is consistent with previous studies, the variation of $n(\mathrm{LAE})/n(\mathrm{LBG})$ within the field of view for each $140\,\mathrm{pMpc}^2$ is found to be as large as a factor of three.
This may suggest a spatially inhomogeneous topology of reionization, but it also leaves open the possibility that the variation is based on the inherent large-scale structure of the galaxy distribution.
Based on the simulations, it may be difficult to distinguish between the two from the current survey.
We also {find} that LAEs in the high LAE density region are more populate high $\mathrm{EW}_0$, supporting that the observed $n(\mathrm{LAE})/n(\mathrm{LBG})$ is more or less driven by the neutral fraction, though the statistical significance is not high.

\end{abstract}

\section{Introduction} \label{sec:introduction}
The reionization process is the transition of the intergalactic medium (IGM) from a neutral to an ionized state.
One of the important features of reionization is the spatial inhomogeneity,
whose topology is closely related to the distributions of the ionizing sources, escape fraction of ionizing photons, the clumpiness of the IGM, and the ionizing photon energy, and so on.
Strong radiation from the first stars and galaxies ionized their surrounding neutral intergalactic medium, creating a sphere of ionized atomic hydrogen, an H\textsc{ii} bubble \citep[e.g.,][]{McQuinn16}. 
The multiplication and progressively larger size of these H\textsc{ii} bubbles resulted in the reionization of nearly all of the available neutral primordial gas, completing the process of cosmic reionization around $z\sim6$. 
Therefore, we expect spatially inhomogeneous distribution of the neutral and ionized IGM, whose characteristic physical scales and topology depend on the histories of clustering and luminosity of the reionizing sources.
{There is an active debate over whether faint galaxies \citep[e.g.,][]{Finkelstein19} or bright \citep[e.g.,][]{Naidu20} galaxies should play a more significant role in the reionization photon budget, but the reionization topology will vary greatly depending on which one contributes more.}
Previous studies have reported some observational evidence of spatially inhomogeneous reionization, primarily from the observations of large scatter of IGM transmission on the background quasars \citep[e.g.,][]{Fan06, Becker15, Bosman18, Yang20}.
Recent observations continue to provide indirect evidence of large H\textsc{ii} bubbles \citep[e.g.,][]{Zheng17, Castellano18, Tilvi20, Meyer20, Hu21}.

The Ly$\alpha$ fraction $X_\mathrm{Ly\alpha}$ is one of the widely used methods for assessing the neutrality of the IGM at the epoch of cosmic reionization (e.g., \citealp{Stark10, Ono12, Curtis-Lake12, Schenker14, Tilvi14, Cassata15}; {see also a review by \citealp{Ouchi20})}.
These studies derive $X_\mathrm{Ly\alpha}$ as the fraction of Lyman break galaxies (LBGs) {that show Ly$\alpha$ emittion lines} at $4 < z < 9$, suggesting the trend that the Ly$\alpha$ fraction abruptly drops from
$z = 6$ to $z = 7$ in contrast to its monotonic increase from $z=4$ to $z=6$.
This can be caused by resonant scattering by intervening neutral hydrogen in IGM during the reionization epoch.
{In addition to} the decrease in the Ly$\alpha$ fraction, the decline in Ly$\alpha$ equivalent width (EW) at $z>6$ has been observed in previous studies, suggesting the suppression of Ly$\alpha$ visibility \citep[e.g.,][]{Treu12, Treu13, Hoag19, Mason19, Jung20}.
However, identifying the Ly$\alpha$ emission lines generally requires spectroscopic observations of LBGs, which is too expensive to enlarge the sample size and survey area.
{Previous studies have limited the survey area to a maximum of about $0.1\,\mathrm{deg}^2$ and the number of targets to about 70 {\citep{Stark11, Schenker14}}.}
Therefore, these studies are unable to depict the spatial variation of the neutral fraction $x_\mathrm{HI}$, and only put the constraints on the averaged value over the survey volume.

To overcome this difficulty, we use a dataset of large-area {($\sim1.5\,\mathrm{deg}^2$)} imagings from Subaru/Hyper Suprime-Cam (HSC; \citet{Furusawa18, Kawanomoto18, Komiyama18, Miyazaki18}).
The large field of view (FoV) of HSC enables us to efficiently detect {Lyman alpha emitters (LAEs)} and LBGs and to illustrate the spatial inhomogeneity of reionization.
In contrast to the previous studies, we detect LAEs based only on the photometric observations and construct LBG and LAE samples separately.
{Though the limiting luminosity of the LAE and LBG samples is not necessarily the same, w}e use a ratio of the number density of LAEs to that of LBGs $n(\mathrm{LAE})/n(\mathrm{LBG})$ as a proxy for the Ly$\alpha$ fraction.
{The same selection criteria as in the observation is applied to the model to consider the difference in limiting magnitude between the two populations when generating the model prediction of $x_\mathrm{HI}$ from $n(\mathrm{LAE})/n(\mathrm{LBG})$.}
One key importance here is that LBGs and LAEs must be simultaneously detected at almost the same redshift to measure $n(\mathrm{LAE})/n(\mathrm{LBG})$.
For this purpose, we have installed a new intermediate-band filter, IB945{, whose central wavelength is $9462\,\mathrm{\AA}$ with {a full width at half maximum (FWHM)} of $330\,\mathrm{\AA}$}.
The IB945 can detect LBGs at $z\sim6.6$, whose redshift range {($\Delta z=0.4$)} is narrower than {that of} {the }typical Lyman break selection {($\Delta z=1.0$)}.
With the wide FoV and the new filter, HSC is the only instrument in the world capable of carrying out this study.

In this paper, we use the imaging data taken by HSC Subaru Strategic Program {\citep[SSP;][]{Aihara18}}\footnote{\url{https://hsc.mtk.nao.ac.jp/ssp/}} and Cosmic HydrOgen Reionization Unveiled with Subaru (CHORUS) project \citep{Inoue20}.
This paper is constructed as follows.
Section \ref{sec:data} describes the observational data we use.
Sample selection is explained in Section \ref{sec:selection}.
We show the distribution of the LAEs and LBGs and draw {the} $n(\mathrm{LAE})/n(\mathrm{LBG})$ map in Section \ref{sec:results}.
Section \ref{sec:discussion} describes the reionization simulation and discusses the neutral fraction and its spatial variation.
We summarize this paper in Section \ref{sec:summary}.

Throughout this paper, we use the AB magnitude system \citep{Oke83}.
We adopt a $\Lambda$CDM cosmology with $h={0.7}$, $\Omega_\mathrm{M}=0.3$, and $\Omega_\Lambda=0.7$, which is consistent with the recent \textit{Planck} observations \citep{Planck20}.

\section{Observational data} \label{sec:data}
\begin{figure}
    \centering
    \includegraphics[]{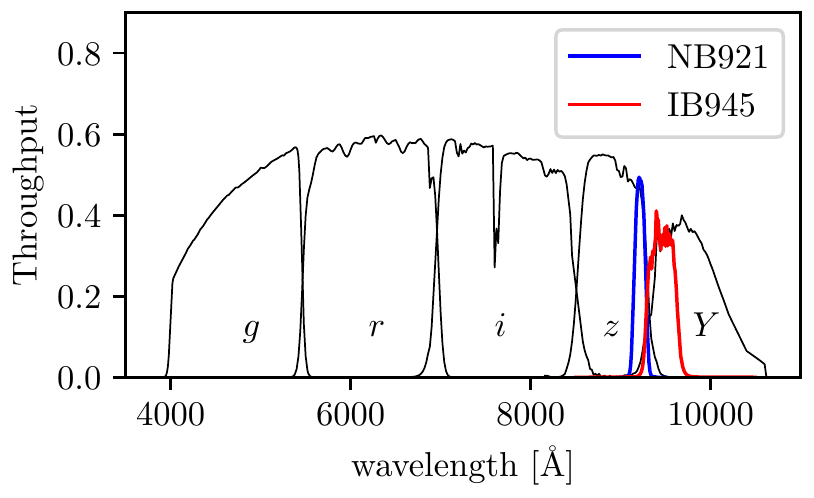}
    \caption{Filter transmission curves of the BB, NB, and IB filters. The blue and red lines show the NB921 and IB945 filters, respectively. The black solid lines represent the $g$-, $r$-, $i$-, $z$-, and $y$-band filters {from left to right}.}\label{fig:transmission}
\end{figure}

We use the HSC-SSP S18A UltraDeep internal data release product of $g$, $r$, $i$, $z$, and $Y$ broad-band (BB), and NB921 images taken from 2014 to 2018.
The details of the SSP observations are described in \citet{Aihara18}.
We refer to \citet{Aihara18b, Aihara19} for the data analysis and the catalog construction.
The NB921 \citep{Ouchi18} has a central wavelength of $9215\,\mathrm{\AA}$ and a FWHM of $135\,\mathrm{\AA}$.
Figure \ref{fig:transmission} shows the transmission curves of the filters.

\begin{figure}
    \centering
    \includegraphics[]{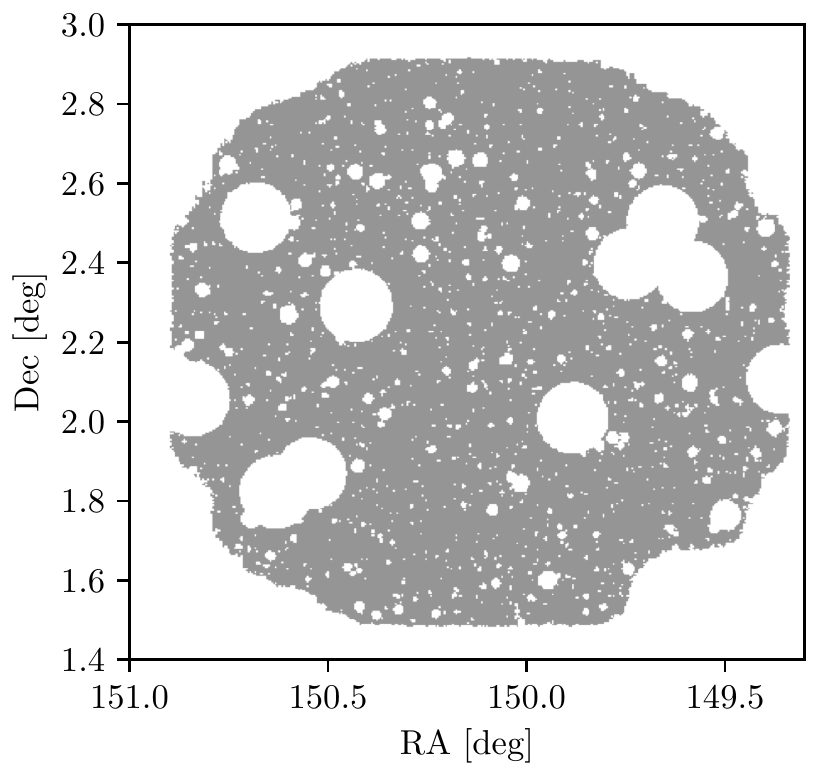}
    \caption{The effective survey area of our data. The mask regions that are combined in all band images are shown in white.}\label{fig:area}
\end{figure}

\begin{figure*}
    \centering
    \includegraphics{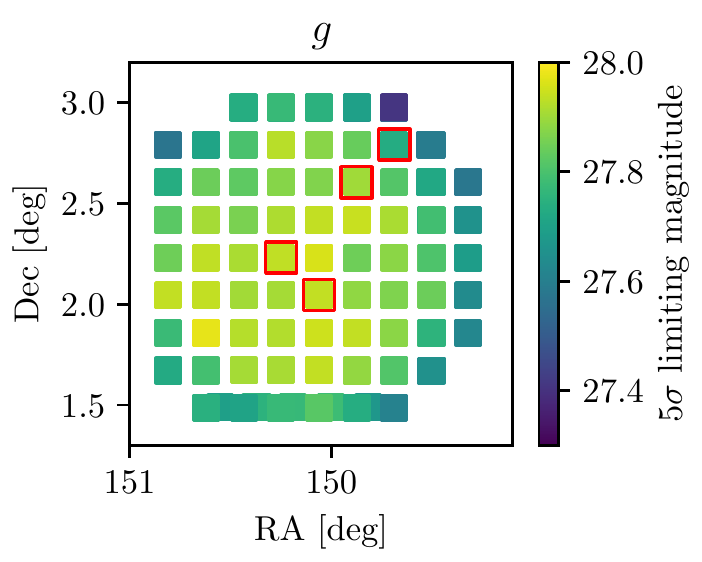}\includegraphics{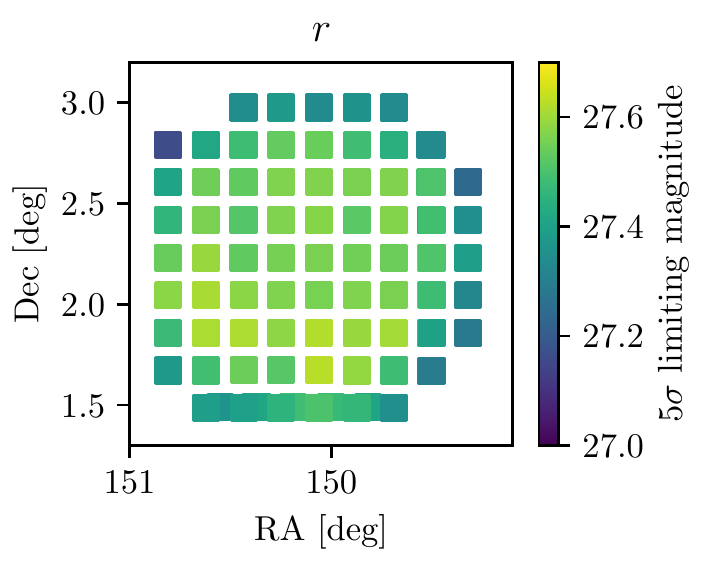}
    \includegraphics{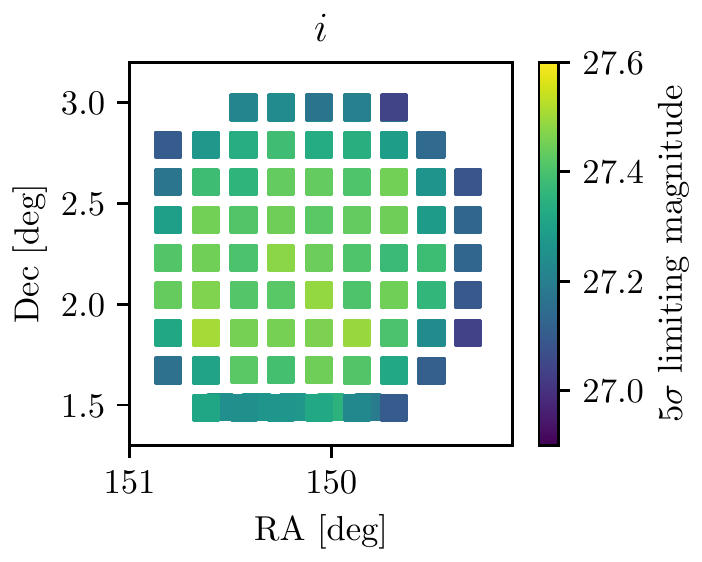}\includegraphics{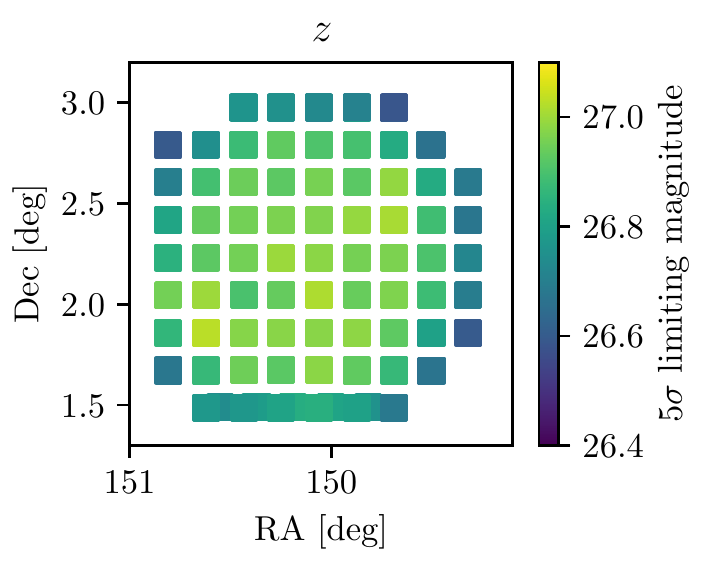}
    \includegraphics{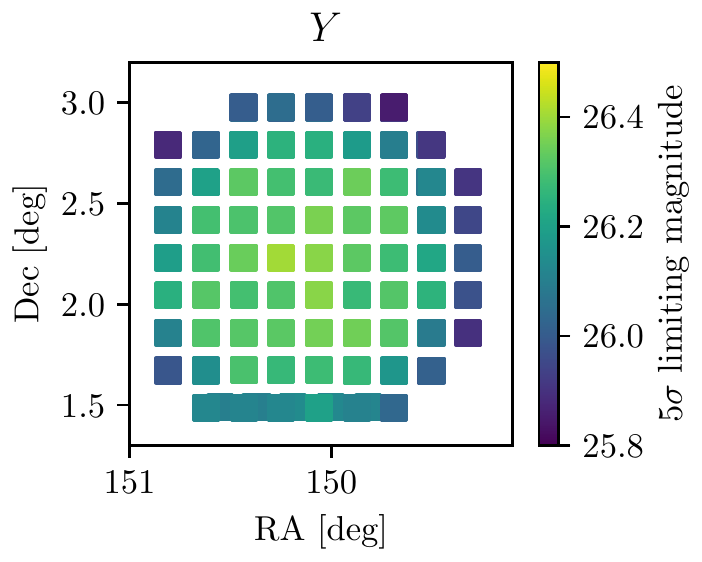}\includegraphics{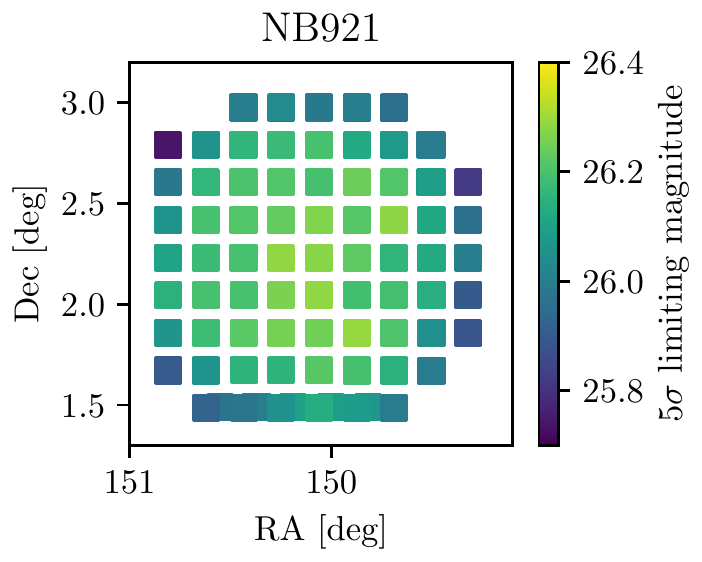}
    \includegraphics{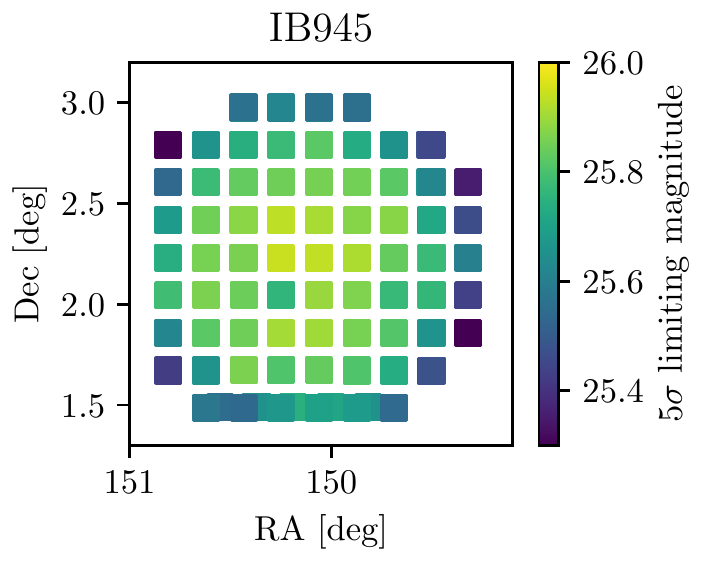}
    \caption{The 5$\sigma$ limiting magnitudes in the $1^{\prime\prime}\!\!.5$ aperture and their spatial variation. {The patches are overlapping due to the definition of the \texttt{tracts} and \texttt{patches} around $\mathrm{DEC}=1.5^\circ$.} {The four red squares show the \texttt{patches} (4,4), (5,3), (6,6), and (7,7) from left to right.}}\label{fig:limitmag}
\end{figure*}

The IB945 images are taken through the CHORUS project \citep{Inoue20}.
We have carried out deep imaging observation with IB945 in the COSMOS field, which is one of the UltraDeep layers of the HSC-SSP.
The HSC data are reduced with HSC pipeline version 6.7 \citep{Bosch18}, which is based on Large Synoptic Survey Telescope (LSST) pipeline \citep{Juric17, Ivezic19, Bosch19}.
The astrometry and photometric calibration is performed based on the data from Panoramic Survey Telescope and Rapid Response System 1 \citep[Pan-STARRS1;][]{Chambers16, Schlafly12, Tonry12, Magnier13}.
The final seeing size is $0^{\prime\prime}\!\!.66$.
The survey area is divided into \texttt{tracts}, and each \texttt{tract} is divided into $9\times9$ \texttt{patches} \citep{Aihara18}.
Table \ref{table:image} summarizes the detail{s} of the imaging data.

\begin{deluxetable}{ccc}
    \caption{Image properties.}\label{table:image}
    \tablehead{
        Filter & 5$\sigma$ limiting magnitude\tablenotemark{a}
         & PSF [${}^{\prime\prime}$]\tablenotemark{b}
    }
    \startdata
    $g$ & 27.9 & 0.84 \\
    $r$ & 27.6 & 0.74 \\
    $i$ & 27.5 & 0.70 \\
    $z$ & 27.0 & 0.65 \\
    $Y$ & 26.4 & 0.77 \\
    NB921 & 26.3 & 0.68 \\
    IB945 & 25.9 & 0.66 \\
    \enddata
    \tablenotetext{a}{{The} 5$\sigma$ limiting magnitude with a $1^{\prime\prime}\!\!.5$ aperture in the \texttt{patch} (4,4).}
    \tablenotetext{b}{PSF size averaged in the entire FoV.}
\end{deluxetable}

In our analysis, we use \texttt{forced} catalogs, in which photometry is measured at the fixed position in all band images.
We use \texttt{convolvedflux} to measure the magnitude after smoothing as a source has the same PSF sizes in all the filters.
The aperture diameter and the PSF size are $1^{\prime\prime}\!\!.5$ and $0^{\prime\prime}\!\!.84$, respectively.

We use \texttt{hscPipe} parameters and flags to exclude objects affected by saturated or bad pixels, and nearby bright stars.
To ensure the images are sufficiently deep, we limit the survey field by using the \texttt{countinput} parameter.
The \texttt{brightobject} mask is automatically generated in the pipeline \citep{Coupon18}.
Because the \texttt{brightobject} mask is not enough around the brightest stars, we visually check the images and add mask regions where the image is affected by the bright objects.
The effective survey area of the IB945 image is $1.46\,\mathrm{deg}^2$ as shown in Figure \ref{fig:area}.
Table \ref{table:flag} indicates the detail of the flags and parameters.

The limiting magnitude in each \texttt{patch} is calculated in the same way as \citet{Inoue20}.
We distribute $\sim 5000$ apertures with $1^{\prime\prime}\!\!.5$ diameter per \texttt{patch} avoiding the masked regions and detected sources and derive the limiting magnitude by fitting the flux distribution with a Gaussian function.
Sky subtraction is performed by a median in an annulus of $1^{\prime\prime}\!\!.0$ width and $2^{\prime\prime}\!\!.5$ inner diameter around the aperture.
The spatial variations in the $1^{\prime\prime}\!\!.5$ aperture limiting magnitudes are shown in Figure \ref{fig:limitmag}.

\section{Selection}\label{sec:selection} 

\subsection{LAE selection}\label{sec:LAEselection}
\begin{figure}
    \centering
    \includegraphics[]{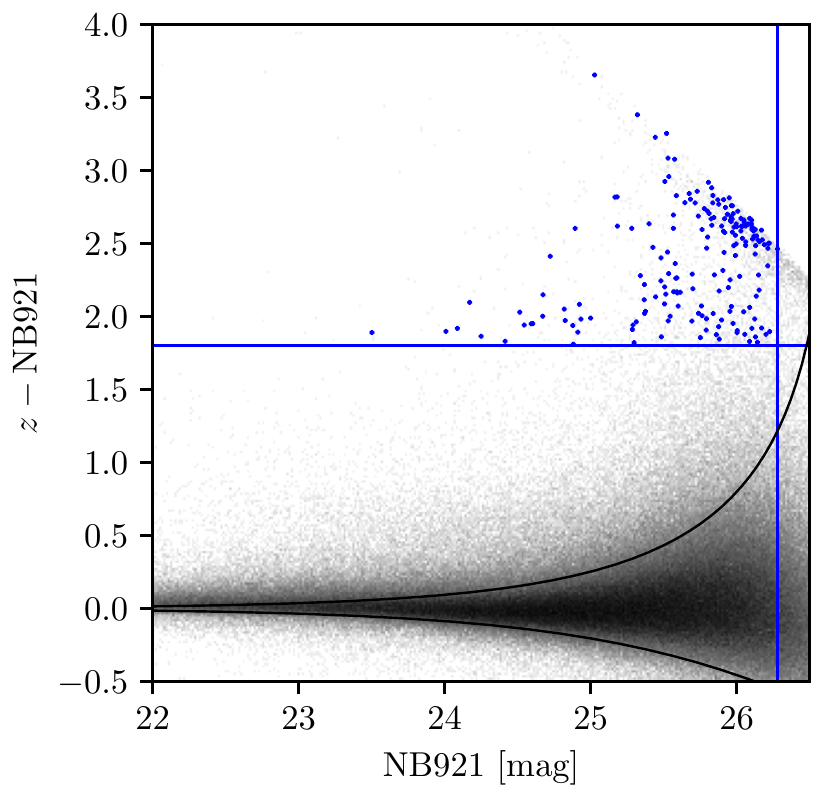}
    \caption{The color magnitude diagram of the LAE candidates. The blue points represent the LAE candidates. The horizontal blue line indicates the color criterion, $z - \mathrm{NB}921 = 1.8$. The vertical blue line shows the limiting magnitude in the \texttt{patch} (4,4). The black solid curve shows the $3\sigma$ error of the $z - \mathrm{NB921}$ color given by {eqation} (\ref{eq:znberr}). The grey dots represent all objects detected in NB921.}\label{fig:cmagdiag}
\end{figure}

We construct {a} $z\sim6.6$ LAE catalog using the HSC data.
The selection of LAEs is based on flux excess in NB compared to adjacent BB filters.
We apply the color selection criteria similar to \citet{Shibuya18}:
\begin{align}
    &\mathrm{NB921} < \mathrm{NB921}_{5\sigma}\\
    \&\ &z - \mathrm{NB921} > 1.8\\
    \&\ &g > g_{5\sigma}\\
    \&\ &r > r_{5\sigma}\\
    \&\ &((z < z_{3\sigma} \ \&\ i - z > 1.3)\nonumber\\
    &\mbox{or}\ z > z_{3\sigma}),
\end{align}
where $\mathrm{NB921}_{5\sigma}$, $g_{5\sigma}$, $r_{5\sigma}$, and $z_{3\sigma}$ represent $5\sigma$ or $3\sigma$ limiting magnitude in each filter.
{To remove low-$z$ contaminants, w}e impose {non-detection} in a shorter wavelength than the Lyman limit and {the existence of} the Lyman break between $i$ and $z$ bands {for sources detected in $z$ band above 3$\sigma$ level}.
The criteria for {non-detection} in $g$ and $r$ bands are relaxed to 5$\sigma$ from $2\sigma$ cut of \citet{Shibuya18}, because the $2\sigma$ cut significantly removes plausible LAE candidates.
{Since the $g$ or $r$ band image is very deep, if a very small 2$\sigma$ level is used as the detection criterion, there are many cases where a small amount of noise is enough to cause the object to be judged as detected.}
{In our dropout sample, detection of Lyman break is guaranteed by relaxing the criterion for non-detection.}
Relaxing the criteria may increase the low-$z$ contaminants, but spectroscopic follow-up observations are necessary for quantitatively evaluate the contamination.
The $z-\mathrm{NB}$ color threshold corresponds to the rest-frame Ly$\alpha$ equivalent width ($\mathrm{EW}_0$) of $\mathrm{EW}_0 > 14\,\mathrm{\AA}$.
{The limiting magnitude in NB921 corresponds to the Ly$\alpha$ luminosity of $2 \times 10^{42}\,\mathrm{erg/s}$.}
We also consider $3\sigma$ error of the $z - \mathrm{NB921}$ color as a function of the NB921 flux $f_\mathrm{NB}$:
\begin{align}
    z - \mathrm{NB921} > -2.5\log(1+3\frac{\sqrt{f_{z,1\sigma}^2 + f_{\mathrm{NB}, 1\sigma}^2}}{f_\mathrm{NB}}), \label{eq:znberr}
\end{align}
where $f_{z,1\sigma}$ and $f_{\mathrm{NB}, 1\sigma}$ are $1\sigma$ flux error in $z$ band and NB921, respectively.
{The }1114 objects are selected by the color-magnitude selection.

\begin{deluxetable*}{lccl}
    \centering
    \caption{The flags and parameters.}\label{table:flag}
    \tablehead{
        Name & Value & Filters & Description
    }
    \startdata
    \texttt{detect\_is\_tract\_inner} & True & - & Source is in the inner region of a \texttt{tract}\\
    \texttt{detect\_is\_patch\_inner} & True & - & Source is in the inner region of a \texttt{patch}\\
    \texttt{countinputs} & $>= 30$ & IB & Number of visits at a position of a source\\
    \texttt{merge\_peak} & True & NB for LAEs & peak detected in a given filter\\
    \texttt{deblend\_nChild} & 0 & $g$, $r$, $i$, $z$, $Y$, NB, and IB & Number of children a source has\\
    \texttt{base\_PixelFlags\_flag\_edge} & False  & $g$, $r$, $i$, $z$, $Y$, NB, and IB & Source is inside exposure region\\
    \texttt{base\_PixelFlags\_flag\_saturatedCenter} & False  & $g$, $r$, $i$, $z$, $Y$, NB, and IB & Source does not have any saturated pixels in the center\\
    \texttt{base\_PixelFlags\_flag\_bad} & False  & $g$, $r$, $i$, $z$, $Y$, NB, and IB & Source does not have any bad pixels in the footprint\\
    \texttt{base\_PixelFlags\_flag\_bright\_object} & False  & $g$, $r$, $i$, $z$, $Y$, NB, and IB & Source footprint does not include pixels\\&&&   \quad affected by bright objects\\
    \texttt{base\_PixelFlags\_flag\_bright\_objectCenter} & False  & $g$, $r$, $i$, $z$, $Y$, NB, and IB & Source center is not affected by bright objects\\
    \texttt{ext\_convolved\_ConvolvedFlux\_1\_4\_5\_flag} & False & $z$, $Y$, and IB for LBGs & Flux measurement fails if true\\ &&$i$, $z$, NB, and IB for LAEs\\
    \enddata
\end{deluxetable*}

We visually inspect the images of LAE candidates to {remove cosmic rays, false detection in NB, and sources that are apparently detected at shorter wavelengths but are considered as non-detection.}
In total, the number of LAE candidates is 189.
{Though the survey area is an order of magnitude larger than the previous studies \citep[e.g.,][]{Stark11, Schenker14}, the number of objects does not increase much as the survey depth is shallower by about one magnitude.}
Figure \ref{fig:cmagdiag} shows the color-magnitude diagram of the LAE candidates.

{The LAE candidates can be contaminated by low-$z$ line emitters with a faint continuum, e.g., H$\alpha$, [O\textsc{iii}], and [O\textsc{ii}] emitters.}
{NB-detected H$\alpha$ emitters, which are non-detected at short wavelengths, must have rest-frame $\mathrm{EW}_0 > 160 \,\mathrm{\AA}$, which is exceptionally large \citep{Fumagalli12}.
To meet the color selection of $z - \mathrm{NB} > 1.8$, the $\mathrm{EW}_0$ of the [O\textsc{ii}] and [O\textsc{iii}] emitters must be greater than $150 \,\mathrm{\AA}$ and $270 \,\mathrm{\AA}$, respectively, but this EW is also extremely large \citep{Reddy18, Khostovan16}.}
{Therefore, we can expect that the contamination of such low-$z$ line emitters is negligible.}
\citet{Shibuya18b} estimate contamination rates in the HSC LAE candidates from spectroscopic confirmations.
The contamination rates are $\simeq 33\%$ and $\simeq 14\%$ for bright ($\mathrm{NB} < 24 \,\mathrm{mag}$) and all LAE candidates, respectively.
{Given that} {we are using the same $z-\mathrm{NB921}$ color cut as theirs}, we expect the contamination rates in our sample are similar{, though different criteria for non-detection of $g$ and $r$ bands may increase the contamination rate}.
\citet{Ono21} construct an LAE sample from the same dataset.
They use $z-\mathrm{NB921} > 1.0$ color cut.
They select LAE candidates with machine learning technique instead of the visual inspection.
{Our LAE catalog includes some fainter objects than theirs due to the difference in the measurement of the limiting magnitude.}
{The number of our LAE candidates that meet their limimting magnitude is 60, and }31 out of the {60} LAE candidates are also included in their LAE catalog.
{Most of the LAE candidates that are not appeared in their catalog are fainter objects.}

\subsection{LBG selection}\label{sec:LBGselection}
\begin{figure}
    \centering
    \includegraphics[]{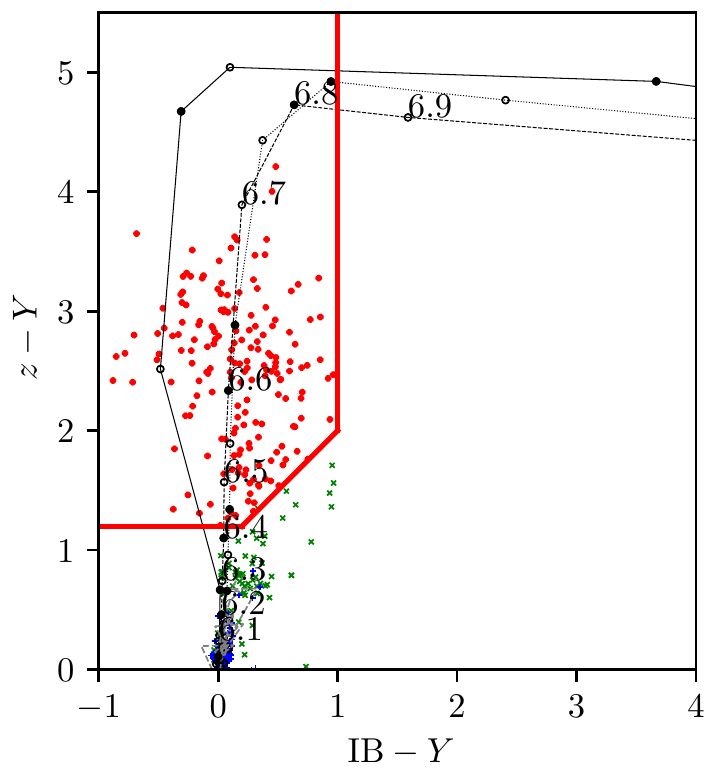}
    \caption{The color-color diagram of the LBG candidates. The red points represent the LBG candidates which pass the visual inspection. For sources fainter than 1$\sigma$ level in the $z$ band, $z$ band magnitudes are replaced with 1$\sigma$ limiting magnitude. The red solid line shows the LBG selection criteria. The black solid, dashed, and dotted lines show the track of LBG model SED constructed by \textsc{Cigale} with $E(B-V)=0.0$, 0.2, and 0.4, respectively. The black circles indicate their redshift from $z=5.5$ to $z=7.0$ with $\Delta z =0.1$ step. The green points indicate dwarf stars taken from \citet{Knapp04}{, and the blue plusses show stars from \citet{Gunn83}}. The dashed grey line shows the typical track of elliptical, Sbc, Scd, and irregular galaxies \citep{Coleman80} from $z=0$ to $2$.}\label{fig:ccdiag}
\end{figure}

We construct a $z\sim6.6$ LBG catalog based on Lyman break color selection technique, which is efficient for selecting sources with a sharp Lyman break and a blue UV continuum.
Selection criteria are as follows:
\begin{align}
    &Y < Y_{5 \sigma}\\
    \&\ &g> g_{5 \sigma}\\ 
    \&\ &r > r_{5 \sigma}\\
    \&\ &z - Y > 1.2\\
    \&\ &\mathrm{IB}945 - Y < 1.0\\
    \&\ &z - \mathrm{IB}945 > 1.0.
\end{align}
To derive these criteria, we use spectral energy distribution (SED) models by \textsc{Cigale} \citep{Boquien19}.
In this modeling, we use the \citet{Bruzual03} single stellar population model and the Salpeter initial mass function \citep{Salpeter55}.
Model parameters are a constant star formation history with an age of 50, 100, 200, and 400 Myr, metallicity of $Z/Z_\odot = 0.2$, the Calzetti extinction law with $E(B-V)= 0.0\mbox{--}0.4$ \citep{Calzetti00}, and the IGM absorption of \citet{Meiksin06}, which is the default IGM model of \textsc{Cigale}.
Figure \ref{fig:ccdiag} shows the evolutionary track of the model SED on the $\mathrm{IB}945 - Y$ vs. $z-Y$ plane.
{The limiting magnitude of the $Y$ band almost corresponds to the rest-frame UV absolute magnitude ($M_\mathrm{UV}$) of $M_\mathrm{UV} < -20.5$ when assuming a flat continuum slope.}
By using IB945, the redshift distribution of LBGs can be made narrower than the typical $z$-dropout selection (see Figure \ref{fig:selecfunccomparison}).
We further explore the redshift distribution of our samples in Section \ref{sec:completeness}.

The 589 objects are selected from the color selection.
We carry out visual inspections to exclude {false detection in the $Y$ band, and sources that are apparently detected at shorter wavelengths but are considered as non-detection.}
Finally, the number of LBG candidates is 179.
{Unlike LAEs, cosmic rays are rarely contaminated, as it requires that they be detected in at least two $Y$ and IB945 bands.}
The color-color diagram of the LBG candidates is shown in Figure \ref{fig:ccdiag}.
{Seven LBG candidates are also included in the LAE catalog, which demonstrates that we detect LBGs at the similar redshift to the LAEs.}
{The reason for the small number of sources selected for both LAE and LBG is that the limiting magnitudes of both samples are not the same. When calculating the neutral fraction from the sample, we take this difference into account and apply these selection criteria to the simulation model.}

{Since the $z-Y$ color of low-$z$ galaxies is small, as shown in Figure \ref{fig:ccdiag}, the contamination rate is expected to be low.}
However, {as in the case of LAE selection, the 5$\sigma$ limiting magnitude is used as a criterion for non-detection in the $g$- and $r$-bands, which may lead to increased contamination.}
{The} quantitative evaluation of the contamination rate is difficult {unless we conduct systematic spectroscopic follow-up} observations of the IB945-selected galaxies.
{\citet{Harikane21} detect 27 $z\sim7$ LBGs in the same field by using a $z-Y$ color criterion.}
{Eighteen of their samples are also included in our LBG sample.}
{Our LBG catalog contains more LBGs than their sample because of the differences in the color selection criteria and the measurement of the limiting magnitude.}

\subsection{Completeness}\label{sec:completeness}
\begin{figure*}
    \centering
    \includegraphics[]{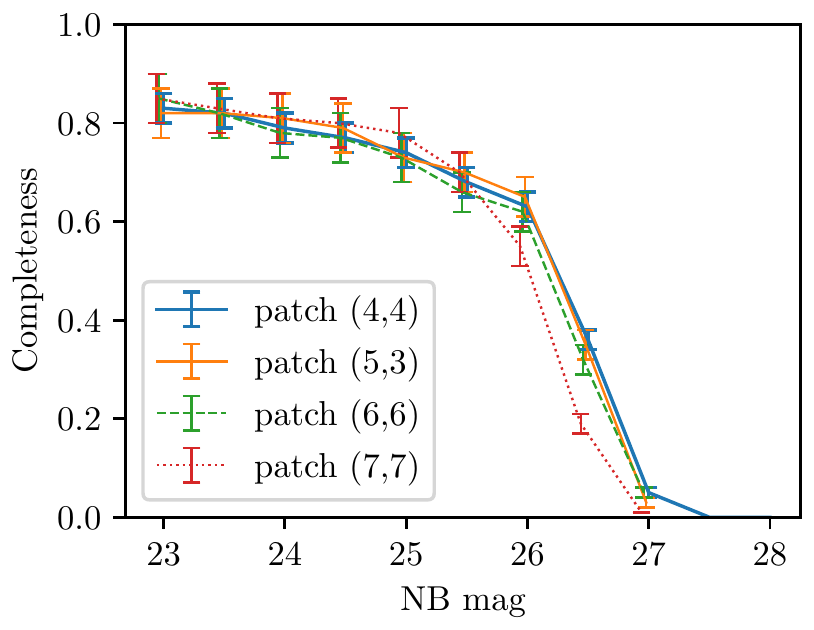}\includegraphics[]{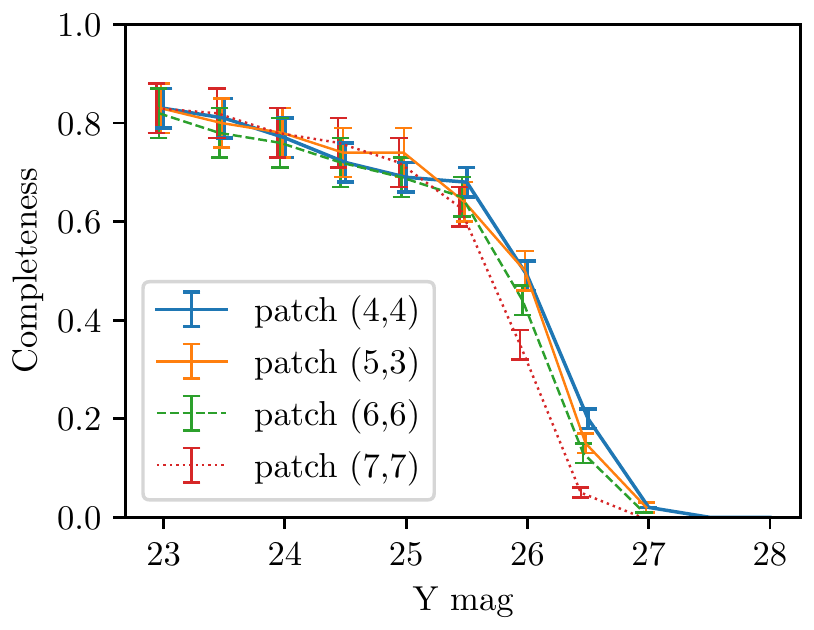}
    \caption{Detection completeness of the NB921 (left) and $Y$ band (right) as a function of magnitude. Thick solid, thin solid, dashed, and dotted lines represent the results in \texttt{patches} (4,4), (5,3), (6,6), and (7,7), respectively. Data points are slightly shifted horizontally for clarity.}\label{fig:detcomp}
\end{figure*}

\begin{figure*}
    \centering
    \includegraphics[]{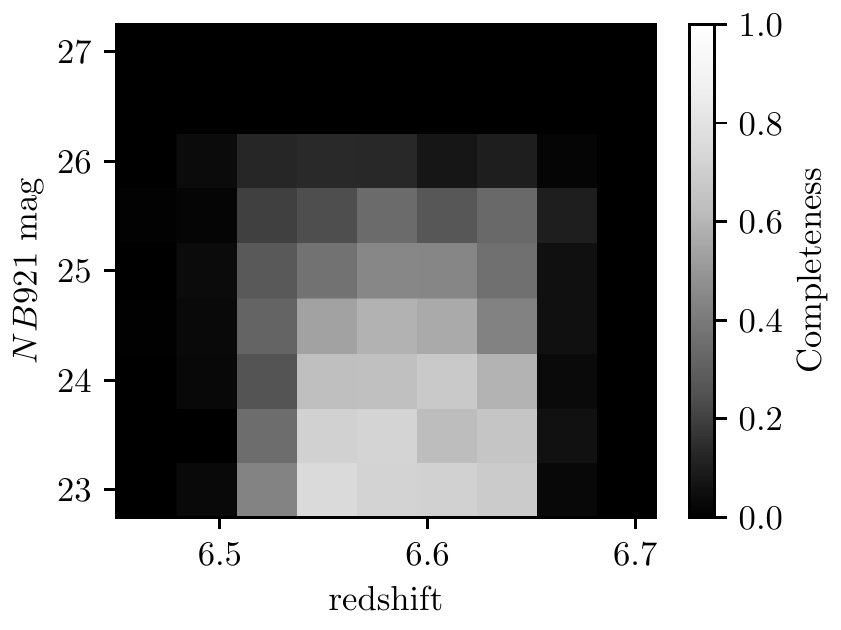}\includegraphics[]{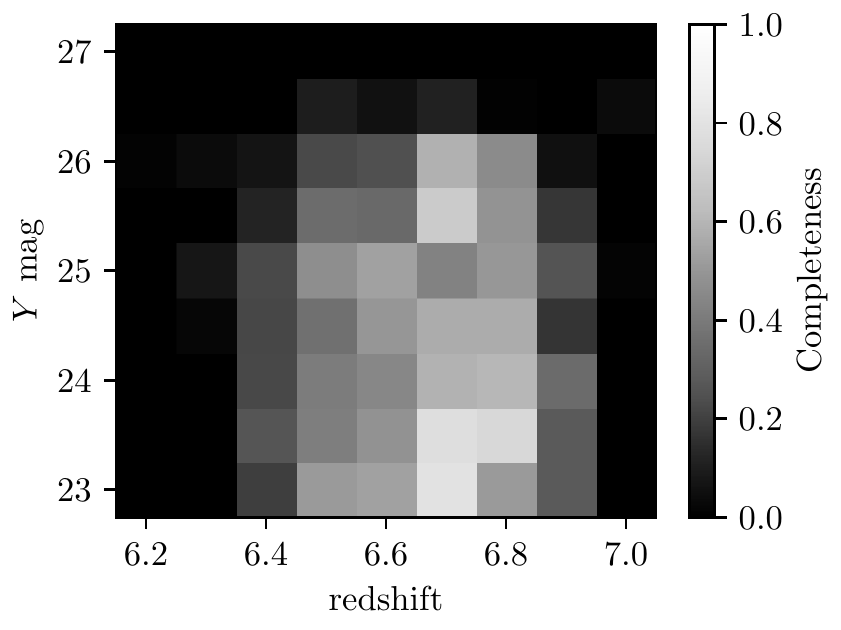} 
    \caption{Selection completeness of our LAEs (left) and LBGs (right) as a function of magnitude and redshift in the \texttt{patch} (4,4).}\label{fig:seleccomp}
\end{figure*}

\begin{figure*}
    \centering
    \includegraphics[]{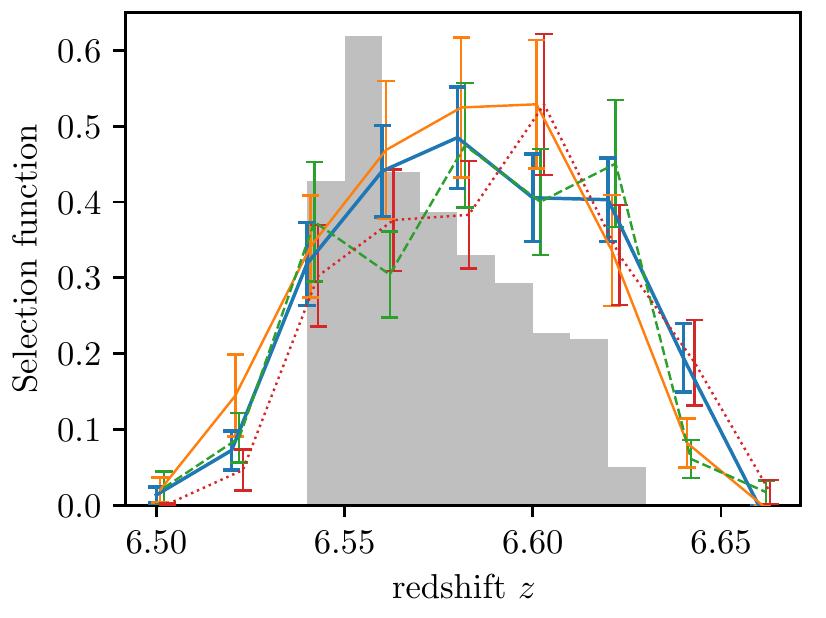} \includegraphics[]{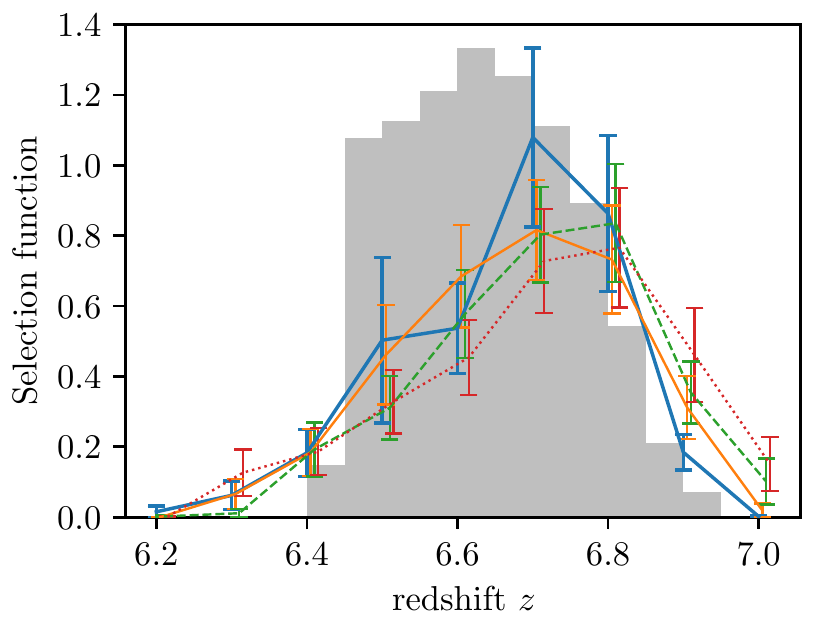}
    \caption{The selection function of the LAEs (left) and the LBGs (right). Colors indicate the \texttt{patches} in the same as Figure \ref{fig:detcomp}. Grey histograms show the redshift distribution of {model} LAEs and LBGs selected from the simulation box (Section \ref{sec:simulation}). {The amplitudes of the histograms are scaled for clarity.}}\label{fig:selecfunc}
\end{figure*}

We estimate the detection and selection completeness of our samples by the following Monte Carlo simulation.
{Here, the detection completeness is defined as the fraction of mock galaxies that are detected in the detection band,
while the selection completeness is defined as the fraction of mock galaxies that satisfy the color selection criteria,
which also takes detection completeness into account.}
To generate a mock LBG catalog, we use the same SED model as described in Section \ref{sec:LBGselection}.
In contrast to the mock LBG catalog, we generate the SED model of LAEs with a simple assumption of a $\delta$-function shaped Ly$\alpha$ emission line, a flat ($f_\nu=\mbox{const.}$) UV continuum, and the IGM attenuation of \citet{Madau95}.
The mock LAEs have the EW of $10 < \mathrm{EW}_0\, [\mathrm{\AA}] < 200$ following the EW distribution obtained from \citet{Shibuya18}.
The redshift of the mock LAEs ranges from 6.50 to 6.66 with an interval of $\Delta z = 0.02$.

We randomly distribute the mock galaxies on the $g$, $r$, $i$, $z$, and $Y$ band, NB921, and IB945 images using \textsc{Balrog} \citep{Suchyta16}.
We consider the total magnitude of the mock galaxies in the range of $23 \leq m \leq 27$ with a 0.5 magnitude step in $Y$ band and NB921 for LBGs and LAEs, respectively.
We assume that the S\'ersic profile with an index of 1.5 and a half-light radius of $0.4\,\mathrm{pkpc}$, which are consistent with \citet{Shibuya15, Shibuya19} at $z\sim6.6$.
The \textsc{Balrog} makes use of \textsc{GalSim} \citep{Rowe15} to simulate the profile of the objects and convolve it with the PSF size.
The PSF model is constructed with \textsc{PSFEx} \citep{Bertin11} from cutout images at each detection processed with \textsc{SExtractor} \citep{bertin96}. 
We detect and measure the mock galaxies using \texttt{hscPipe} version 6.7 in the same way as described in Section \ref{sec:data}.

The detection completeness is defined as a number fraction of {the mock galaxies that are} successfully detected in the detection band images ($Y$ and NB921 for LBGs and LAEs, respectively).
We regard sources that are detected within $0^{\prime\prime}\!\!.5$ from {the positions of the mock galaxies} and with a difference of 0.5 magnitude or less from the input magnitude as successfully detected objects \citep{Inoue20}.
We apply the same selection to the mock galaxies as described in Sections \ref{sec:LAEselection} and \ref{sec:LBGselection} and calculate the selection completeness of our LAE and LBG samples.
The selection completeness is defined as a number ratio of the mock galaxies satisfying the selection criteria to {all} the mock galaxies.
{The LAE selection function considers $z-\mathrm{NB}$ color.
Since the wavelength ranges of filter transmission of $z$ band and $\mathrm{NB}921$ overlap, emission lines detected in NB do not necessarily satisfy the $z-\mathrm{NB}$ color selection.
It is important to consider the selection completeness for LAEs.}
The uncertainty of the completeness takes into account the Poisson errors of the numbers of the embedded and recovered mock galaxies.

As shown in Figure \ref{fig:limitmag}, the spatial variation of the limiting magnitude in the survey area is approximately circularly symmetric and the limiting magnitude becomes gradually shallow towards the edge of the FoV.
We divide the survey area into four regions with respect to the distance from the center and assume that the completeness is constant in each region.
We use the \texttt{patches} (4,4), (5,3), (6,6), and (7,7), which are 0, 16, 32, and 48 arcmin away from the center of the FoV {(Figure \ref{fig:limitmag})}.
We carry out this simulation in the four \texttt{patches} to estimate the spatial variation of completeness in the FoV.
Figure \ref{fig:detcomp} shows the detection completeness for LBGs and LAEs.
This figure shows slightly lower completeness at the edge of the FoV.
The detection completeness is less than unity even on the bright end.
This is because the embedded mock galaxies overlap with the bright objects in the image.
Figure \ref{fig:seleccomp} shows the selection completeness as a function of magnitude and redshift for LBGs and LAEs in the \texttt{patch} (4,4).

\begin{figure}
    \centering
    \includegraphics[]{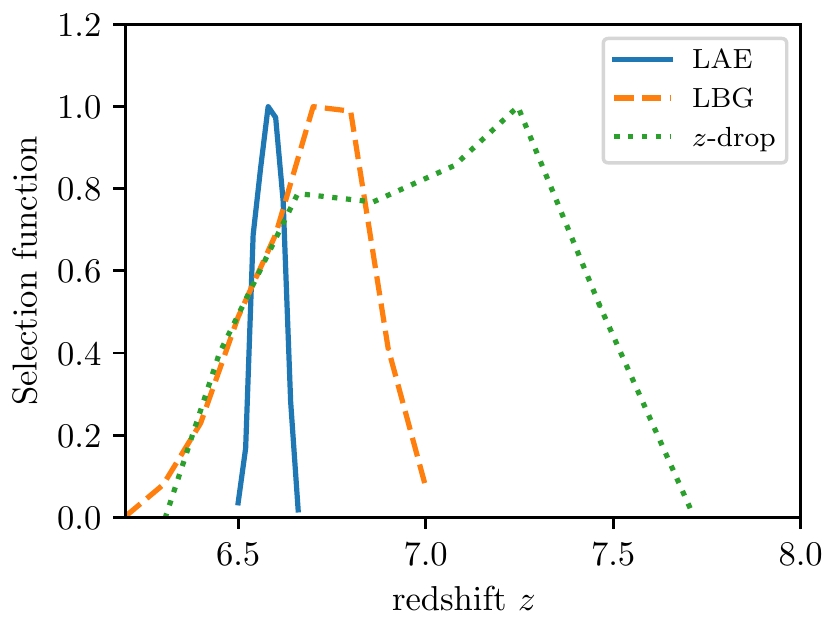}
    \caption{Normalized selection functions of the samples. The blue solid and the orange dashed line represent the LAEs and LBGs, respectively. We also show the selection function of $z$-dropout galaxies from \citet{Ono18} with the green dotted line for comparison.}\label{fig:selecfunccomparison}
\end{figure}

In Section \ref{sec:fLya_xHI_relation}, we sample model galaxies in the same redshift range as the observation to make comparisons.
For this purpose, we derive the redshift distribution without taking detection incompleteness into account, which we call the selection function.
We define the selection function as the selection completeness divided by the detection completeness.
In each patch, we derive the selection function for each magnitude.
The amplitude of the selection function of LAE strongly depends on the NB magnitude because the fainter objects have larger uncertainty in the observed color, while the selection function of LBG has almost no changes in terms of the magnitude.
We average the selection functions over the magnitude weighted with the observed number count.
The error of the selection function is calculated from the uncertainty of the detection and selection completeness.
Figure \ref{fig:selecfunc} shows the selection function of LAEs and LBGs.
We do not find any systematic variations of the selection function in the four \texttt{patches}.
The selection function should be less than or equal to unity, but there is one point that is greater than unity in the LBG selection function, which is just caused by the calculation.
{The selection completeness is calculated in each magnitude and redshift, while the detection completeness is averaged over redshift.
Thus, the selection completeness is not necessarily smaller than the detection completeness.}
We also show in Figure \ref{fig:selecfunc} the redshift distribution of model galaxies (Section \ref{sec:simulation}) selected with the same color selection criteria.
The redshift ranges are similar to the selection functions while the peaks of the distribution are shifted to lower redshift.
{This is because higher redshift galaxies are more difficult to detect due to their fainter magnitude.}
In Figure \ref{fig:selecfunccomparison}, we compare the selection function of the samples.
While the width of the LBG selection function is broader than that of LAEs, it is narrower than that of the typical $z$-dropout selection.
This confirms that IB945 allows us to detect LBGs at similar redshift to LAEs.
The remaining difference in the selection function between LBG and LAE is taken into account in the model galax{ies} to be compared, along with the difference in the limiting magnitude, so as to have exactly the same selection function with the observation (Section \ref{sec:fLya_xHI_relation}).

\section{Results} \label{sec:results}

\subsection{Surface density}\label{sec:surfacedensity}
\begin{figure}
    \centering
    \includegraphics[]{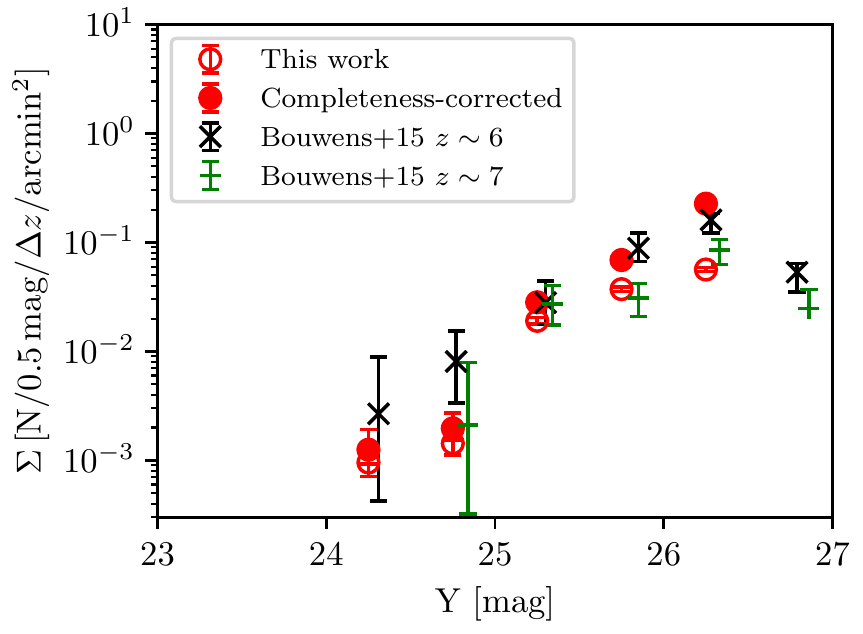}
    \caption{Surface number density of LBGs at $z \sim 6.6$ as a function of $Y$ band magnitude. Filled and open circles represent the raw and {the detection} completeness corrected surface densities of our $z \sim 6.6$ LBGs ($\Delta z = 0.4$), respectively. Black crosses and green plusses indicate the surface densities of $z\sim6$ ($\Delta z=0.8$) and $7$ ($\Delta z=1.0$) galaxies of \citet{Bouwens15}, respectively.}\label{fig:countLBG}
\end{figure}

\begin{figure}
    \centering
    \includegraphics[]{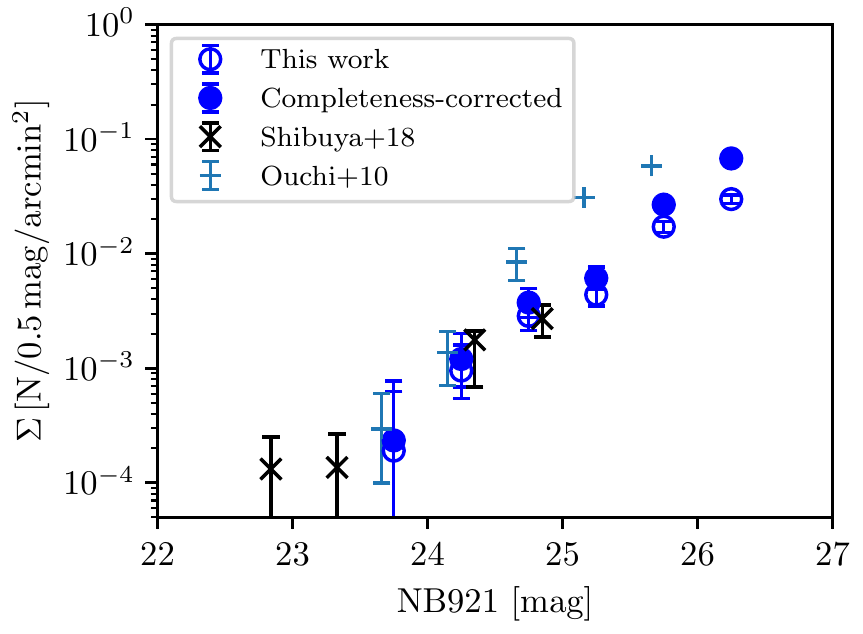}
    \caption{Surface density of LAEs at $z = 6.6$ as a function of NB921 magnitude. Filled and open circles represent the raw and {the detection} completeness corrected surface densities of our $z \sim 6.6$ LAEs, respectively. For comparison, we also plot results of \citet{Shibuya18} and \citet{Ouchi10}. The points are slightly shifted horizontally for clarity.}\label{fig:countLAE}
\end{figure}

Applying the selection criteria described in Section \ref{sec:selection}, we obtained 179 LBGs and 189 LAEs at $z \sim 6.6$.
Figures \ref{fig:countLBG} and \ref{fig:countLAE} show the surface densities of LBGs and LAEs as a function of magnitude, respectively.
Since the redshift range ($\Delta z$) of our LBG selection is different from that of \citet{Bouwens15}, we correct it by the width of the selection function estimated in Section \ref{sec:completeness}.
The surface densities of LBGs and LAEs corrected by the detection completeness are found to be consistent with the previous studies.
{The surface density of LAEs is lower than that of \citet{Ouchi10}, though this is because the \texttt{MAG\_AUTO} of \textsc{SExtractor}, which they use as the total magnitude, is biased towards brighter magnitudes.}
By integrating the surface density {corrected by the detection completeness} down to the limiting magnitudes, we obtain $n(\mathrm{LAE}) / n(\mathrm{LBG}) = 0.76 \pm 0.10$ as an average of the entire FoV.\footnote{Note that this value has a different definition from the so-called Ly$\alpha$ fraction, which has been measured so far. Throughout this paper, we use the term $n(\mathrm{LAE})/n(\mathrm{LBG})$ to distinguish it from the Ly$\alpha$ fraction. See section \ref{sec:comparison} for a comparison of the two.}
In this calculation, we only consider the Poisson error and the error of the completeness as the uncertainty of $n(\mathrm{LAE})/n(\mathrm{LBG})$.

\begin{figure}
    \centering
    \includegraphics[]{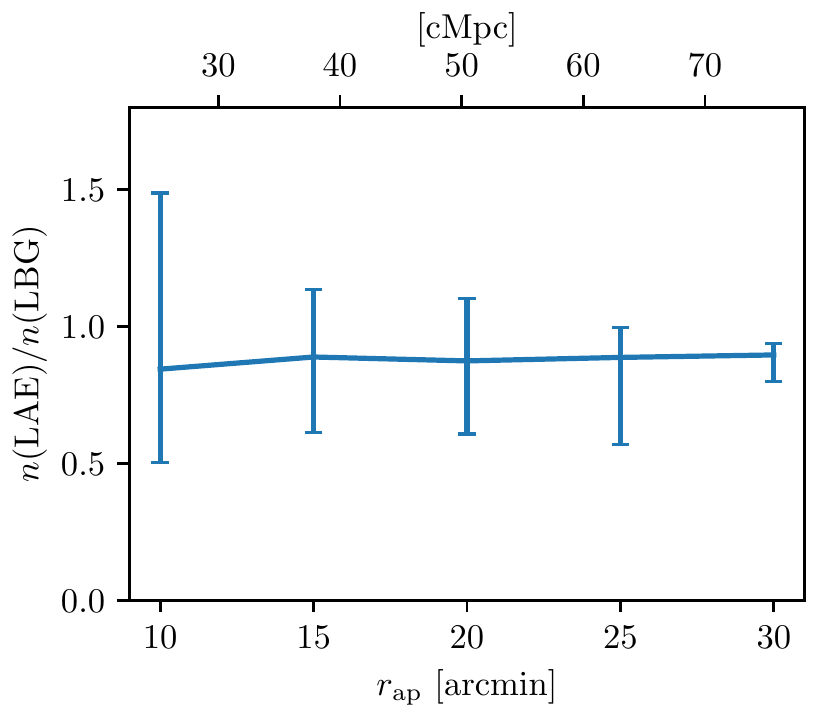}
    \caption{The median and 68 percentile range of $n(\mathrm{LAE})/n(\mathrm{LBG})$ in the FoV as a function of aperture size.}\label{fig:nratioscatter_obs}
\end{figure}

To evaluate the scatter of $n(\mathrm{LAE})/n(\mathrm{LBG})$, we distribute $\sim5000$ apertures of $\mathrm{10}$--$30$ arcmin radii on the survey area and count the number of LAEs and LBGs within them.
We exclude apertures where the masked region{s} cover more than 20\% of the apertures.
Figure \ref{fig:nratioscatter_obs} shows the median and scatter of $n(\mathrm{LAE})/n(\mathrm{LBG})$ as a function of aperture radii.
There is deviation from the value calculated for the entire FoV ($n(\mathrm{LAE}) / n(\mathrm{LBG}) = 0.76 \pm 0.10$) even with $30^\prime$ aperture{s}.
This is because the number ratio in the entire FoV is derived by integrating the number density down to the limiting magnitude, which is a little different from counting the numbers in the apertures.
{Both approaches would give consistent results, but the difference in procedure of counting and completeness correction may cause slightly different results.}

To see the spatial variation of $n(\mathrm{LAE})/n(\mathrm{LBG})$ over the field, we use $20^\prime$ ($\sim 50\,\mathrm{cMpc}$) radius aperture in the following analysis.
{In general, using a small aperture ($r_\mathrm{ap}<15^\prime$) will result in a larger variation of $n(\mathrm{LAE})/n(\mathrm{LBG})$, as expected, because fewer objects are contained in the aperture.}
{On the other hand, an aperture of $>25^\prime$ radius, which is too large compared to the FoV of HSC, is not appropriate for the study of spatial inhomogeneity.}
With the $20^\prime$ radius, the field averaged $n(\mathrm{LAE}) / n(\mathrm{LBG})$ is $0.84 ^{+0.23}_{-0.27}$.
Since this calculation takes the spatial variation of $n(\mathrm{LAE})/n(\mathrm{LBG})$ {between the apertures} into account, the uncertainty is larger than that described above.
The difference between the maximum and minimum values is as large as a factor of three.
In the following, we use this value as our result of $n(\mathrm{LAE})/n(\mathrm{LBG})$.

\subsection{Spatial distribution} \label{sec:distribution}
\begin{figure*}
    \centering
    \includegraphics[]{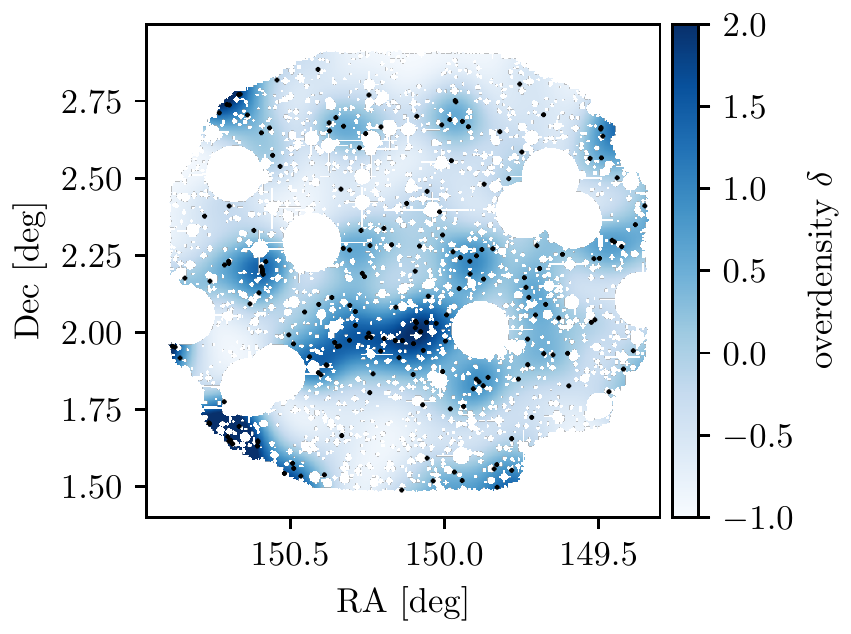}\includegraphics[]{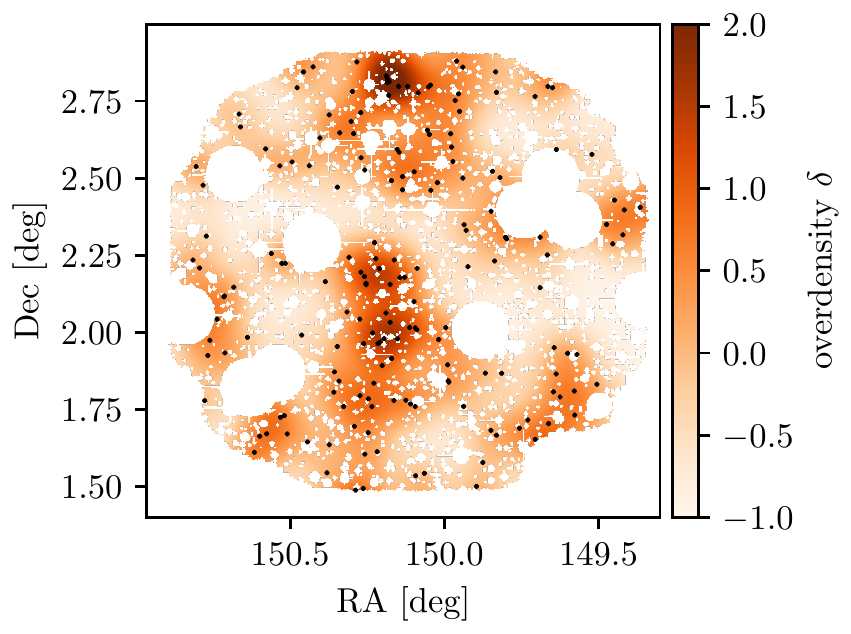}
    \caption{Boundary corrected density distribution of the LAEs (left) and LBGs (right). Black dots indicate the distribution of candidates of the LAEs and LBGs. Color bar represents the overdensity $\delta = (\sigma - \bar{\sigma}) / \bar{\sigma}$. Masked regions are shown in white (see Figure \ref{fig:area}).}\label{fig:density}
\end{figure*}

We calculate the density distribution using the Gaussian kernel density estimation.
The surface density at a coordinate $X = (\mathrm{RA},\,\mathrm{Dec})$ is given as
\begin{align}
    \sigma(X) = \sum_i K(X, X_i),
\end{align}
where $K$ is the kernel function and $X_i = (\mathrm{RA}_i,\,\mathrm{Dec}_i)$ is the coordinate of the $i$-th galaxy.
We choose a Gaussian as the kernel function:
\begin{align}
    K(X, X_i) = \frac1{2\pi\lambda^2} \exp\left(-\frac{{r_i}^2}{2\lambda^2}\right),
\end{align}
where $r_i$ is the projected separation between $X$ and $X_i$, and $\lambda$ is the bandwidth parameter.
The optimal bandwidth is calculated by maximizing the likelihood cross-validation \citep[LCV;][]{Hall82} in the same way as \citet{Chartab20}.
The LCV is defined as
\begin{align}
    \mathrm{LCV}(\lambda) = \frac1N \sum_{k=1}^N \log\sigma_{-k}(X_k),
\end{align}
where $N$ is the total number of the objects and $\sigma_{-k}(X_k)$ is the surface density at the position $(X_k)$ calculated excluding the $k$-th object.
The optimized values are $3^\prime\!.63$ and $3^\prime\!.75$ for LBGs and LAEs, respectively.
We adopt $4^\prime\!.00$ as the common bandwidth for LBGs and LAEs to compare them under the same condition.
The spatial inhomogeneity of the depth is corrected with the detection completeness by assuming the circularly symmetric variation as described in Section \ref{sec:completeness}.

Since the survey area is limited and has the masked regions, correcting the boundary effect is necessary.
The true density distribution is estimated as \citep[see][]{Chartab20, Jones93}:
\begin{align}
    \sigma_\mathrm{true}(X) = \frac{\sigma(X)}{\int_S K(X,\, X_i)\dd S},
\end{align}
where the integration is calculated over the effective area of the survey field ($S$).

We define the overdensity $\delta$ as
\begin{align}
    \delta = \frac{\sigma - \bar{\sigma}}{\bar{\sigma}},
\end{align}
where $\bar{\sigma}$ is the mean density averaged over the entire effective survey area after the completeness correction and the boundary correction.
Figures \ref{fig:density} shows the density distribution of LBGs and LAEs, respectively.
{The LAE overdensity regions located slightly south of the center of the FoV are consistent with the results reported in \citet{Higuchi19} and \citet{Zhang20}.}

\begin{figure}
    \centering
    \includegraphics[]{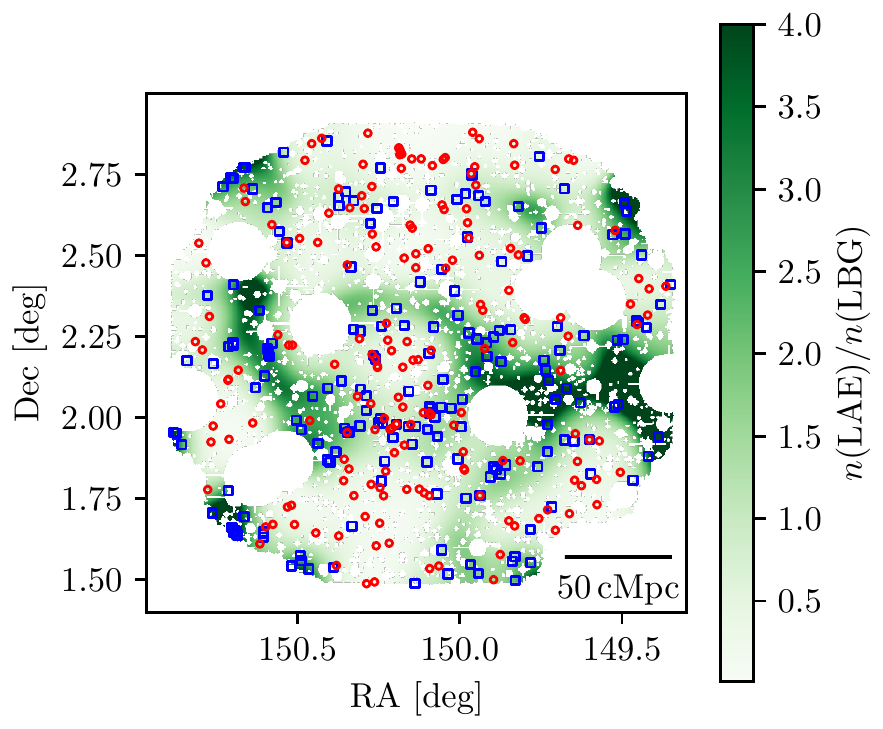}
    \caption{The map of $n(\mathrm{LAE})/n(\mathrm{LBG})$. The color map is constructed by dividing the density distribution of LAEs by that of LBGs (Figure \ref{fig:density}). The blue squares and red circles show the distribution of LAEs and LBGs, respectively.}\label{fig:nratiomap}
\end{figure}

Using the whole sample of LBGs and LAEs, we draw the map of $n(\mathrm{LAE})/n(\mathrm{LBG})$.
From Figure \ref{fig:nratiomap}, it is interpreted that the reionization is proceeding in the high LAE density region, while it is delayed in the low LAE density region, though the intrinsic distribution of these galaxy populations must also be taken into account, which {is} discussed in Section \ref{sec:patchy_vs_lss}.

\section{Discussion} \label{sec:discussion}

\subsection{Neutral fraction}\label{sec:neutralfraction}

\subsubsection{Reionization simulation}\label{sec:simulation}

We carry out a reionization simulation to predict the neutral fraction $x_\mathrm{HI}$ from $n(\mathrm{LAE})/n(\mathrm{LBG})$ based on \citet{Inoue18}.
{Simulating star formation and ionizing photon escape requires high-resolution radiative hydrodynamical {(RHD)} simulations.}
{However, an RHD simulation in a large-scale ($>100\,\mathrm{cMpc}$) box is not feasible because of the huge numerical costs.}
{Therefore, we divide the simulation into following two steps.}
First, we conduct a high-resolution RHD simulation to model galaxies and IGM in a $20\,\mathrm{cMpc^3}$ box.
The RHD simulation takes into account radiative feedback, which regulates star formation.
Then, the recipe constructed in the first step is used for solving the radiative transfer of ionizing photons in a large-scale $N$-body simulation{ in a box of $(162\,\mathrm{cMpc})^3$}.
\citet{Inoue18} present several models {in terms of the production rate and the escape fraction of Ly$\alpha$ photons} to simulate LAE SED{.}
{They} find that the Model G {reproduces} the observed luminosity function, auto correlation function, and LAE fraction; therefore, we adopt Model G.
{In the Model G, the Ly$\alpha$ photon production rate is determined as a function of halo mass without any fluctuations.}
{The Ly$\alpha$ escape fraction in the Model G depends on halo mass with a fluctuation to model the stochasticity in the transfer of Ly$\alpha$ photons.}
The reionization simulation assumes two models, the Late and Mid reionization models.
The mean neutral fractions of the Late and Mid reionization models at $z\sim6.6$ are $0.4$ and $0.0$, respectively.
The detail of this simulation is described in \citet{Inoue18}.
We generate 20 (10 for the Late reionization and the other 10 for Mid reionization) light cone outputs from $z=5.5$ to $z=9$ by randomly shifting and rotating the snapshots of the simulation box.
Each light cone output has a $\sim1\,\mathrm{deg}^2$ FoV.

\subsubsection{Correlation between Ly$\alpha$ fraction and neutral fraction}\label{sec:fLya_xHI_relation}
Using the simulation described above, we derive the relation between $x_\mathrm{HI}$ and $n(\mathrm{LAE}) / n(\mathrm{LBG})$. 
We select {model} LAEs and LBGs in the light cones based on exactly the same selection functions (Figure \ref{fig:selecfunc}) as observed in Section \ref{sec:completeness}{ to take into account the difference in the redshift range between the observed LAEs and LBGs}.
The model galaxies are randomly chosen according to the value of the selection function at their redshift.
The model LAEs meet the Ly$\alpha$ luminosity of $L_\mathrm{Ly\alpha} > 2.5\times 10^{42}\,\mathrm{erg\,s^{-1}}$ and $\mathrm{EW_0} > 14\,\mathrm{\AA}$, and the model LBGs have the rest-frame UV absolute magnitude of $M_\mathrm{UV} < -20.3$.
These conditions are similar to those estimated from the selection criteria described in Sections \ref{sec:LAEselection} and \ref{sec:LBGselection}.
While we use the two reionization models, the neutral fractions are concentrated around $x_\mathrm{HI}=0.4$ and $0.0$ at $z=6.6$.
To obtain the continuous relation over a wide range of $x_\mathrm{HI}$,
we shift the selection function along the redshift axis with $\Delta z = 0.1$ steps in a range of $6<z<8$ without changing the shape, and we sample the model LAEs and LBGs at each redshift.
We assume a little evolution of both galaxy populations over $6<z<8$.
We count $n(\mathrm{LAE})$ and $n(\mathrm{LBG})$ in randomly distributed apertures with radii of $20$ arcmin ($\sim 50\,\mathrm{cMpc}$), then calculate $n(\mathrm{LAE}) / n(\mathrm{LBG})$.
The neutral fraction is calculated in each aperture; averaged over the volume of the aperture with $20^\prime$ radius and $\Delta z = 0.1$ ($\sim40\,\mathrm{cMpc}$) at the peak of the selection functions.
Because the variation from different light cones is small, we combine all the 20 light cone outputs.
{In general, the Mid reionization model shows higher redshift than the Late reionization model at the same $x_\mathrm{HI}$.}
{Despite the difference in redshift, the both reionization models show similar results of $x_\mathrm{HI}$--$n(\mathrm{LAE}) / n(\mathrm{LBG})$ relation; therefore the effect of redshift evolution on the relation is considered to be negligible over $6<z<8$.}

\begin{figure}
    \centering
    \includegraphics[width=3.2in]{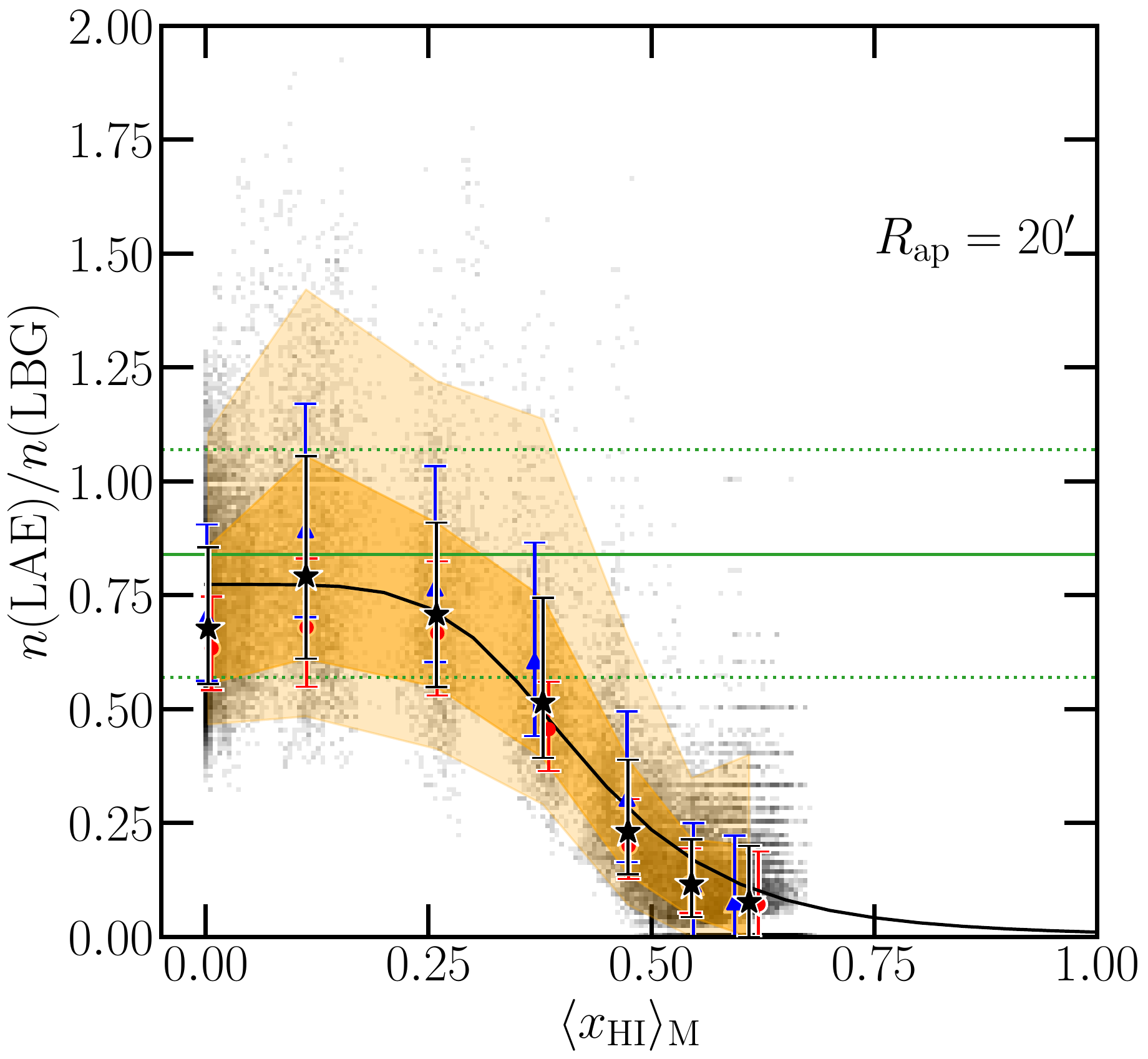}
    \caption{The relation between $n(\mathrm{LAE}) / n(\mathrm{LBG})$ and the neutral fraction $x_\mathrm{HI}$. The background grey colors show the all data points calculated in the light cone outputs, and the red circles and blue triangles show their median and 16th--84th percentile range for the Late and Mid reionization models, respectively. The black stars take into account both reionization models. The black curve and the orange shade indicate the median and the 68 and 95 percentile ranges of this relation. The {horizontal solid and dashed green lines show the median and the $1\sigma$ scatter of} the obserbed $n(\mathrm{LAE}) / n(\mathrm{LBG})${, respectively}.}\label{fig:nratio_xHI_relation}
\end{figure}

We show in Figure \ref{fig:nratio_xHI_relation} the relation between $n(\mathrm{LAE}) / n(\mathrm{LBG})$ and $x_\mathrm{HI}$.
The $n(\mathrm{LAE}) / n(\mathrm{LBG})$ is inversely correlated with $x_\mathrm{HI}$ though the relation flattens out in the range of $x_\mathrm{HI}<0.25$ because most LAEs reside in ionized bubbles and their Ly$\alpha$ lines are less sensitive to the neutral gas in the highly ionized universe.

\subsubsection{Field averaged neutral fraction and comparisons with previous studies}\label{sec:comparison}

In Section \ref{sec:surfacedensity}, we obtain $n(\mathrm{LAE})/n(\mathrm{LBG}) = 0.84 ^{+0.23}_{-0.27}$ as an average of the entire FoV.
{By} using the relation shown in Figure \ref{fig:nratio_xHI_relation}, we estimate the neutral fraction averaged over the survey field as $x_\mathrm{HI} < 0.4$ {by the 1$\sigma$ range of $n(\mathrm{LAE}) / n(\mathrm{LBG})$}.
In addition to the large scatter in $n(\mathrm{LAE}) / n(\mathrm{LBG})$ in the simulation data, observed scatter in $n(\mathrm{LAE}) / n(\mathrm{LBG})$ is large due to the shallow depth of the observation.
The constraint for $x_\mathrm{HI}$ is only given as an upper limit.
Deeper observations are required to more strictly constrain $x_\mathrm{HI}$.

Since the faintest objects, the completeness of which is low ($\sim0.5$), have a dominant contribution to the total number of objects, this could be prone to uncertainty.
To address this problem, we use the objects that are 0.5 mag brighter than the limiting magnitude and derive $x_\mathrm{HI}$ in the same way as described above.
When calculating the $n(\mathrm{LAE}) / n(\mathrm{LBG})$--$x_\mathrm{HI}$ relation {from the simulation model}, we {also} use 0.5 mag brighter selection criteria for the model LBGs.
We raise the lower limit of the Ly$\alpha$ luminosity of the model LAE selection by $\times 1.6$ (corresponding to 0.5 mag bright) since in the model, LAE selection is not based on photometric magnitude.
{The numbers of the model LBGs and LAEs decrease by factors of 0.3 and 0.4 with the 0.5 mag brighter selection criteria, respectively.}
{This is consistent with the surface density of the observed LBGs and LAEs (Figures \ref{fig:countLBG} and \ref{fig:countLAE}).}
The same result is obtained in this analysis; therefore, the $x_\mathrm{HI}$ estimate is found to be robust.
In Appendix \ref{append:nratio_xhi_relation}, we describe the other selection procedures to further explore the robustness of our results.

\begin{figure}
    \centering
    \includegraphics[]{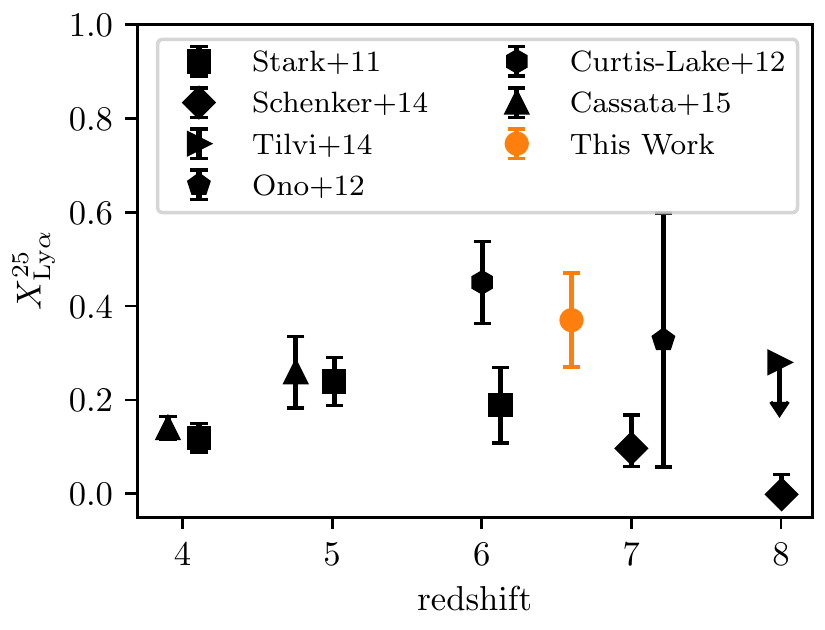}
    \caption{Ly$\alpha$ fraction $X_\mathrm{Ly\alpha}^{25}$ as a function of redshift. The previous results are taken from \citet{Stark10, Schenker14, Tilvi14, Ono12, Curtis-Lake12, Cassata15}.}\label{fig:lyafraction}
\end{figure}

In previous studies, the Ly$\alpha$ fraction $X_\mathrm{Ly\alpha}$ is defined as the fraction of LBGs that show a Ly$\alpha$ emission line.
These studies spectroscopically confirm Ly$\alpha$ emission lines among their photometric LBG sample.
On the other hand, the redshift range and the limiting magnitude of our LAEs and LBGs are {not exactly the same} since we detect them separately.
To compare our result with the Ly$\alpha$ fraction estimates of the previous studies $X_\mathrm{Ly\alpha}^{25}$, 
{we additionally constrain LAE as detected in $Y$ band with $5\sigma$ significance ($Y < Y_{5\sigma}$).}
{The $Y$ band magnitude is a proxy for the rest-frame UV continuum flux, and $Y_{5\sigma}$ corresponds to $M_\mathrm{UV} < -20.3$, which is similar to the depth of the LBG sample.}
{We also limit LAEs with $\mathrm{EW}_0 > 25\,\mathrm{\AA}$.}
{The $\mathrm{EW}_0$ is calculated from the photometry in NB921 and the $Y$ band.}
{The number of LAEs that satisfy $Y < Y_{5\sigma}$ and $\mathrm{EW}_0 > 25\,\mathrm{\AA}$ is 15.}
Since the redshift range is different between LBGs and LAEs, we correct the difference by multiplying the ratio of the widths of the selection functions described in Section \ref{sec:completeness}.
Figure \ref{fig:lyafraction} shows the comparison of $X_\mathrm{Ly\alpha}^{25}$ with the previous studies.
Overall, our measurement of the field-averaged $n(\mathrm{LAE})/n(\mathrm{LBG})$ is consistent with the $X_\mathrm{Ly\alpha}^{25}$ estimates of the previous studies within the error.

\subsubsection{Spatial inhomogeneity of neutral fraction in the simulation box}\label{sec:inhomogeneity}
\begin{figure}
    \centering
    \includegraphics[]{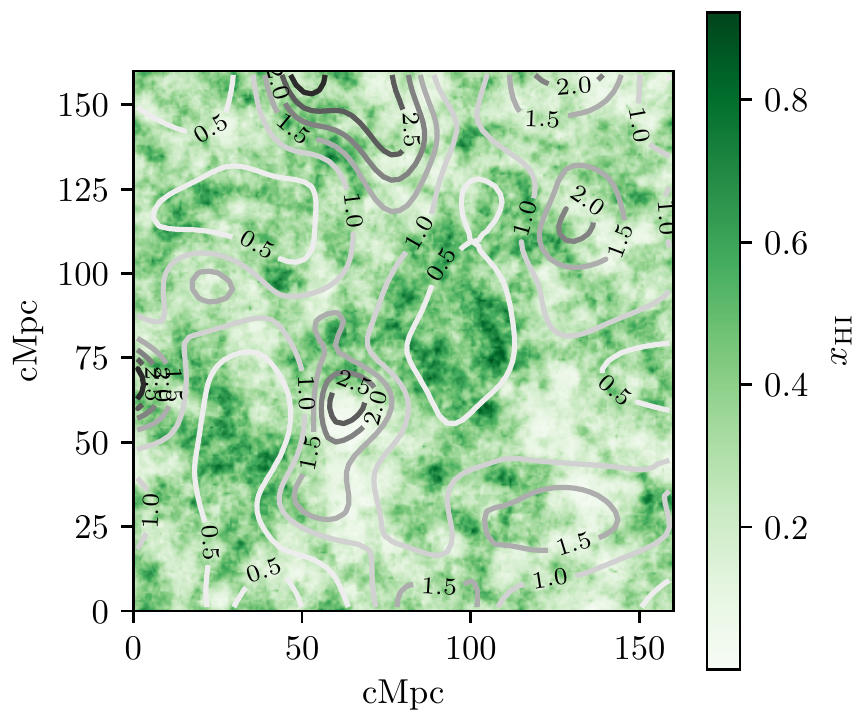}
    \caption{$x_\mathrm{HI}$ map and $n(\mathrm{LAE})/n(\mathrm{LBG})$ distribution in a simulation box of the Late reionization model. The green color shows the $x_\mathrm{HI}$ and the contours indicate the $n(\mathrm{LAE})/n(\mathrm{LBG})$.}\label{fig:nratio_xhi_map}
\end{figure}

\begin{figure}
    \centering
    \includegraphics[]{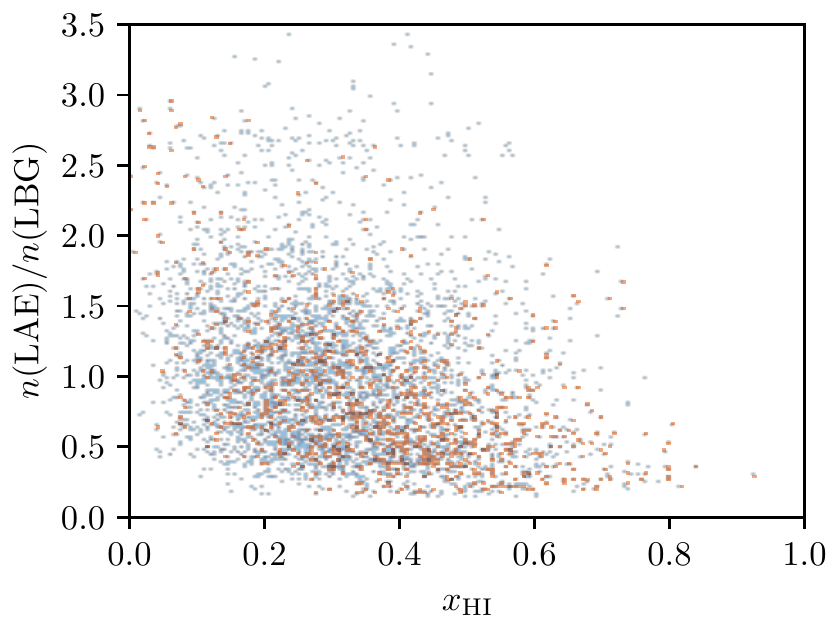}
    \caption{The relation between $x_\mathrm{HI}$ and the $n(\mathrm{LAE})/n(\mathrm{LBG})$ in the simulation box when dividing the map in Figure \ref{fig:nratio_xhi_map} into a grid with $1^\prime$ ($2.5\,\mathrm{cMpc}$) side. The blue points show whole region shown in the map, and the orange points are from the central region with a square of $80\,\mathrm{cMpc}$ per side (25\% of the total area of the simulation box).}\label{fig:nratio_xhi_map_relation}
\end{figure}

As shown in Figure \ref{fig:nratiomap}, we find the spatial variation of $n(\mathrm{LAE}) / n(\mathrm{LBG})$.
The $n(\mathrm{LAE}) / n(\mathrm{LBG})$ map could be converted to $x_\mathrm{HI}$ map using Figure \ref{fig:nratio_xHI_relation} to see a possible spatial inhomogeneity in the reionization process.
However, since the two are almost uncorrelated for $x_\mathrm{HI}<0.25$, and the relation has a large scatter, it may be difficult to draw $x_\mathrm{HI}$ map from the observed data{.}
{T}hus, we will use only the simulated data to see the spatial correlation between $n(\mathrm{LAE}) / n(\mathrm{LBG})$ and $x_\mathrm{HI}$.
We average $x_\mathrm{HI}$ in a redshift slice of $6.54 < z < 6.63$.
The selections of LAEs and LBGs are based on the color and magnitude in the same as the observation (Section \ref{sec:selection}), and we calculate $n(\mathrm{LAE}) / n(\mathrm{LBG})$ in the same way as described in Section \ref{sec:distribution}.
Figure \ref{fig:nratio_xhi_map} shows the $x_\mathrm{HI}$ and $n(\mathrm{LAE}) / n(\mathrm{LBG})$ map at $z=6.6$ in one of the light cone models.
We find a moderate correlation (correlation coefficient $r=-0.25$ and the p-value $p < 0.001$) between $x_\mathrm{HI}$ and $n(\mathrm{LAE}) / n(\mathrm{LBG})$ when we divide the map into a grid with $1^\prime$ ($2.5\,\mathrm{cMpc}$) side as shown in Figure \ref{fig:nratio_xhi_map_relation}.
{The apertures near the edge of the simulation box have less effective area overlapping with the simulation box, increasing the uncertainty of the relation.}
When we limit to the {central region with a square of $80\,\mathrm{cMpc}$ per side,} where boundary effect is small, the correlation becomes strong ($r=-0.47$ and $p < 0.001$).
The result, at least in the simulation, demonstrates that the spatial variation of $n(\mathrm{LAE}) / n(\mathrm{LBG})$ seen in Figure \ref{fig:nratiomap} implies the patchy reionization topology, i.e., the reionization is proceeding in the high LAE density region, while it is delayed in the low LAE density region. 
The high LAE density regions could correspond to the H\textsc{ii} bubbles at $z\sim6.6$.

\subsubsection{Is the observed inhomogeneity caused by patchy reionization or intrinsic LSS?}\label{sec:patchy_vs_lss}
\begin{figure}
    \centering
    \includegraphics[]{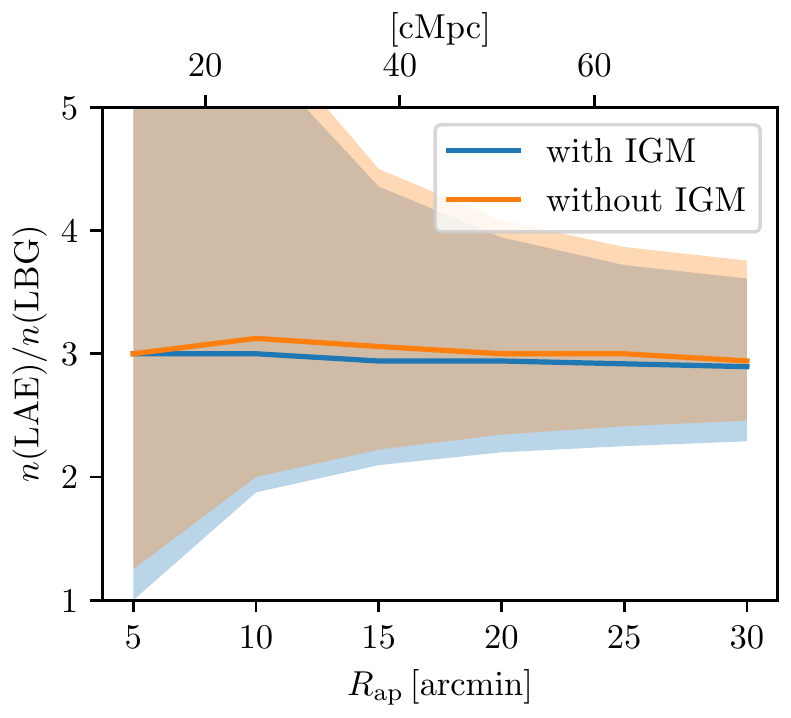}
    \caption{The median and 1$\sigma$ scatter of $n(\mathrm{LAE})/n(\mathrm{LBG})$ as a function of aperture size in the Late reionization model. The blue and orange indicate with and without IGM absorption. {The absolute value of $n(\mathrm{LAE})/n(\mathrm{LBG})$ is different from Figures \ref{fig:nratioscatter_obs} and \ref{fig:nratio_xHI_relation} because LAE and LBG are both selected with the top-hat redshift distribution of $\Delta z=0.1$.}}\label{fig:nratioscatter_sim}
\end{figure}

In the previous section, we discuss a possible reionization topology from our observed spatial variation of $n(\mathrm{LAE})/n(\mathrm{LBG})$, although the number ratio of LAEs and LBGs, $n(\mathrm{LAE}) / n(\mathrm{LBG})$ could depend also on their intrinsic distribution, the large scale structure (LSS).
To distinguish the reionization topology from the LSS, we estimate the intrinsic scatter of $n(\mathrm{LAE}) / n(\mathrm{LBG})$ without the IGM attenuation from the simulation data.
The simulation calculates three kinds of Ly$\alpha$ luminosities; intrinsic, escaped from a halo, and transmitted through IGM.
The third one is the Ly$\alpha$ luminosity with the IGM absorption and the second one is that without the IGM absorption ($x_\mathrm{HI}=0$).
We compare the scatter of $n(\mathrm{LAE}) / n(\mathrm{LBG})$ using the two Ly$\alpha$ luminosities in selecting LAEs.
{The reionization inhomogeneity further increases the scatter of $n(\mathrm{LAE}) / n(\mathrm{LBG})$ in addition to the intrinsic scatter due to the LSS.
Therefore, if we can confirm that the scatter with IGM is larger than that without IGM, we can conclusively distinguish between the two.}
Because in the Mid reionization model ($x_\mathrm{HI} =0.0$ at $z=6.6$) the difference between the two Ly$\alpha$ luminosities is negligible, we use the Late reionization model ($x_\mathrm{HI}=0.4$ at $z=6.6$) in this analysis.

We select LAEs and LBGs in the same as Section \ref{sec:fLya_xHI_relation}, but we adopt the top-hat redshift distribution of $\Delta z=0.1$ for LAEs and LBGs at $z=6.6$ (see Appendix \ref{append:nratio_xhi_relation}).
The number of LBGs does not change with or without IGM absorption, while
the number of LAEs with the IGM absorption is smaller by a factor of 0.6 than that without the IGM absorption{.}
In this comparison, {the number of objects in both cases with and without IGM absorption, should be comparable, because we need to compare the scatter of both cases, as distinct from the difference in Poisson errors.}
{T}herefore, we randomly choose LAEs with the probability of 0.6 in the case of no IGM attenuation.
{In the case of IGM absorption, the Ly$\alpha$ attenuation depends on the neutral fraction, random choosing enables uniform reduction of the number of LAE without being affected by IGM absorption.}
Choosing randomly increases the uncertainty, but its effect is negligibly small.
We distribute $\sim 5000$ apertures on the FoV and calculate $n(\mathrm{LAE}) / n(\mathrm{LBG})$ within them.
When the area overlapping with the simulation box is less than 80\% of the aperture, we exclude the aperture, which degrades the result due to the large Poisson error.
Varying the aperture radius from $5^\prime$ to $30^\prime$ (corresponding to $1.6\mbox{--}10\,\mathrm{pMpc}$), we show in Figure \ref{fig:nratioscatter_sim} the scatter of $n(\mathrm{LAE}) / n(\mathrm{LBG})$ with and without the IGM absorption.
There is no difference between them probably due to the small number of the sample or a low neutral fraction at $z\sim6.6$; therefore we conclude that distinguishing the reionization topology from the LSS may be difficult with this survey depth.

\begin{figure}
    \centering
    \includegraphics[]{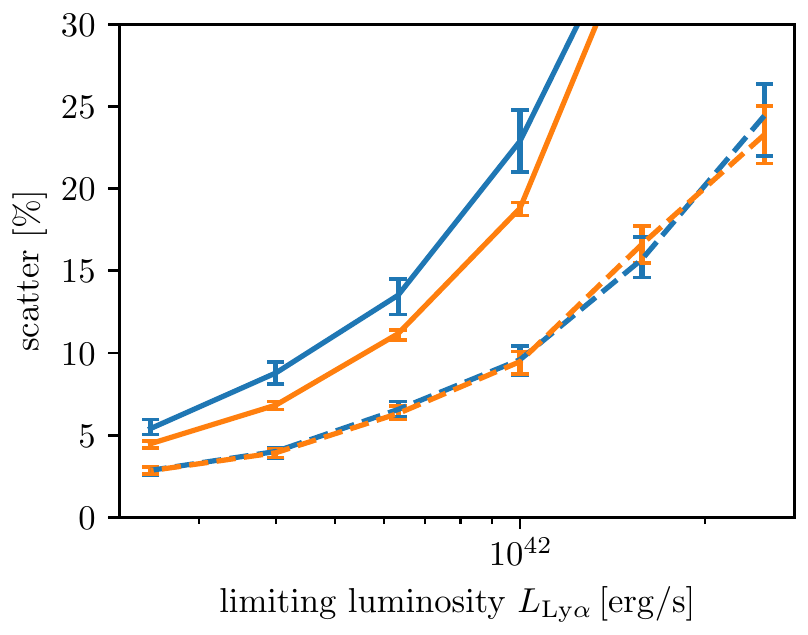}
    \caption{The scatter of $n(\mathrm{LAE})/n(\mathrm{LBG})$ relative to the median value as a function of limiting Ly$\alpha$ luminosity in the Late reionization model. Colors are the same as Figure \ref{fig:nratioscatter_sim}. The dashed (solid) lines show the result of $z=6.6$ ($8.0$). The grey dotted line shows the current survey depth. {The error bars are estimated by bootstrapping the model galaxies before
    calculating $n(\mathrm{LAE})/n(\mathrm{LBG})$ within apertures.}}\label{fig:nratioscatterdeep}
\end{figure}

To investigate how deep observation is required for this aim, we carry out the same analysis including fainter objects in the simulation.
We change the limiting Ly$\alpha$ luminosity down to $L_\mathrm{Ly\alpha} > 2.5\times 10^{41}\,\mathrm{erg/s}$ with $0.2$ dex interval and the limiting rest-frame UV absolute magnitude down to $M_\mathrm{UV} < -17.5$ mag with $0.5$ mag step.
To match the expected number of LAEs with and without the IGM absorption, we randomly choose LAEs without the IGM absorption as described above.
Figure \ref{fig:nratioscatterdeep} shows the scatter of $n(\mathrm{LAE}) / n(\mathrm{LBG})$ within the apertures with $20'$ radius between with and without the IGM absorption as a function of limiting Ly$\alpha$ luminosity.
{The error bars are calculated by bootstrapping the model galaxies before calculating $n(\mathrm{LAE}) / n(\mathrm{LBG})$ within the apertures.}
When we include the fainter objects the scatter of $n(\mathrm{LAE}) / n(\mathrm{LBG})$ becomes smaller.
However, we cannot find significant difference between with and without IGM absorption in the simulation even with two magnitude deeper observation{s (limiting luminosity $L_\mathrm{Ly\alpha} \sim 4\times 10^{41}\,\mathrm{erg/s}$)} than the current survey.

At $z=6.6$, IGM absorption hardly decreases the number of faint objects since the neutral fraction is not so high ($x_\mathrm{HI}\sim0.4$).
That seems to be one of the reasons why we cannot see significant difference between with and without IGM absorption.
We carry out the same analysis at $z=8.0$ ($x_\mathrm{HI}\sim0.8$) to verify that the comparison of $n(\mathrm{LAE})/n(\mathrm{LBG})$ can distinguish reionization topology from the LSS.
Since the number of objects declines at higher redshift, we use $\Delta z=0.4$ at $z=8.0$.
In the same way as $z=6.6$, we derive the scatter of $n(\mathrm{LAE})/n(\mathrm{LBG})$ as a function of limiting Ly$\alpha$ luminosity.
{We might be able to detect} a difference between with and without IGM absorption {seen in  Figure \ref{fig:nratioscatterdeep}} {assuming deep observations (limiting luminosity $L_\mathrm{Ly\alpha} \sim 4\times 10^{41}\,\mathrm{erg/s}$)}
{while the difference is small.}
This suggests that observation at the earlier phase of reionization {may} be able to detect spatially inhomogeneous nature in the universe with a sufficient amount of neutral hydrogen.

The variance of $n(\mathrm{LAE})/n(\mathrm{LBG})$ in the simulations used in this study is so large that there is no clear difference in the scattering with and without IGM absorption, as shown in Figures \ref{fig:nratioscatter_sim} and \ref{fig:nratioscatterdeep}.
This large variance may be due to uncertainties in the model, such as stochastic processes in the LAE model, as well as Poisson noise and the LSS.
When deeper observational data are considered, Poisson noise decreases, but the variance from the LSS and LAE models still exceeds the variance from $x_\mathrm{HI}$ {at $z=6.6$}(Figure \ref{fig:nratioscatterdeep}).
{If we can make this observation deeper at higher-$z$ to study the earlier stage of reionization, we may be able to detect spatial variation of $x_\mathrm{HI}$.}
However, this is only a validation using this simulation, and other simulations based on different reionization scenarios may give different results, and it is not clear if this is the case in the real universe.
Although this simulation model reproduces each observation well in terms of the evolution of {the Ly$\alpha$ luminosity function, the auto correlation function}, and {the} Ly$\alpha$ fraction, it may be beyond the scope of this simulation model to reproduce the variance as well.
Further investigation of the model is necessary.

\subsection{Equivalent Width Distribution}
\begin{figure}
    \centering
    \plotone{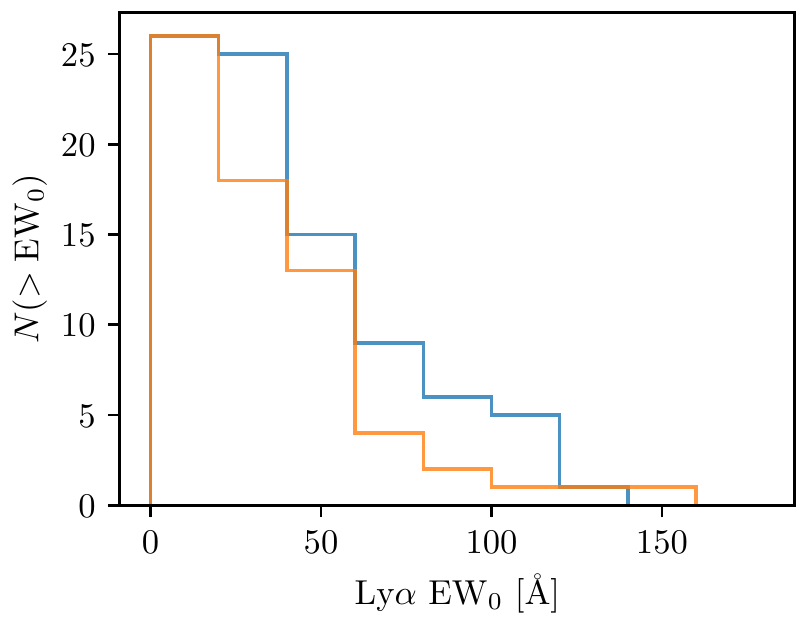}
    \caption{The difference in the Ly$\alpha$ EW distribution of the LAE candidates detected in $Y$ band between the high and low LAE density. The blue (orange) histogram indicates the cumulative EW distribution of the LAEs in the top (bottom) quartile of $n(\mathrm{LAE})/n(\mathrm{LBG})$ at their positions.}\label{fig:ew}
\end{figure}

\begin{figure}
    \centering
    \includegraphics[]{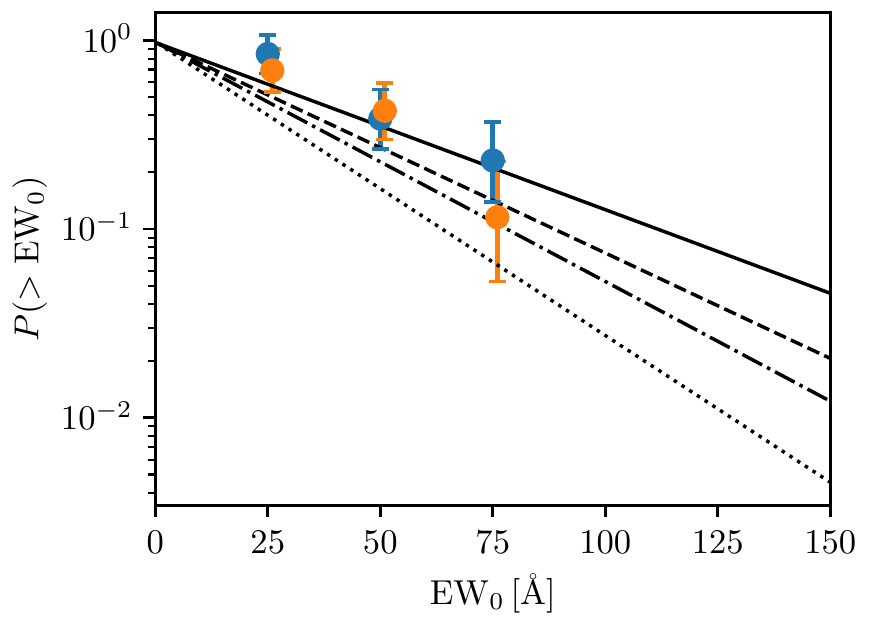}
    \caption{Cumulative probability distribution of $\mathrm{EW}_0$ of the LAEs. Blue (orange) marks indicate the high (low) $n(\mathrm{LAE})/n(\mathrm{LBG})$ environment. Black curves show the model prediction of \citep{Dijkstra11} in the universe with $x_\mathrm{HI}=0$, $0.23$, $0.51$, and $0.74$ from top to bottom.}\label{fig:ew_xhi}
\end{figure}

The rest-frame Ly$\alpha$ EW distribution also could be a plausible probe of reionization \citep{Treu12, Treu13, Jung20}.
We calculate the $\mathrm{EW}_0$ of the LAEs in the same way as \citet{Shibuya18}.
In this analysis, we exclude sources undetected in $Y$-band, which would cause large uncertainties on the $\mathrm{EW}_0$ measurements.
The $\mathrm{EW}_0$ distribution of our LAE sample is similar to \citet{Shibuya18}.
{The Kolmogorov-Smirnov test does not find any significant difference between the two distributions (p-value $p=0.36$).}
To investigate the relation between the $\mathrm{EW}_0$ and the neutrality, we divide the sample into high and low LAE density regions based on $n(\mathrm{LAE})/n(\mathrm{LBG})$ shown in Figure \ref{fig:nratiomap} at the positions of LAEs.
When we compare the $\mathrm{EW}_0$ distribution between the top and bottom quartiles of $n(\mathrm{LAE})/n(\mathrm{LBG})$, as shown in Figure \ref{fig:ew}, the LAEs in the high LAE density region more populate $\mathrm{EW}_0 > 100\,\mathrm{\AA}$, though the statistical significance is not high.
We compare the $\mathrm{EW}_0$ distribution with the model prediction of \citet{Dijkstra11} as shown in Figure \ref{fig:ew_xhi}.
While the error bar is large, low $n(\mathrm{LAE})/n(\mathrm{LBG})$ environment might indicate a high neutral fraction, implying that the LAEs in the highest $n(\mathrm{LAE})/n(\mathrm{LBG})$ environment are surrounded by ionized bubbles.
It is in line with the relation that increasing the neutral fraction reduces $n(\mathrm{LAE})/n(\mathrm{LBG})$.
While we assume a flat UV continuum when calculating the $\mathrm{EW}_0$, the uncertainty of the UV slope affects the $\mathrm{EW}_0$.
However, considering the uncertainty is difficult because most of our LAEs are detected only in $Y$-band redward of the Ly$\alpha$.
The UV slope might depend on the environment, and it will be investigated by future observations.

\section{Summary}\label{sec:summary}
In this paper, we, for the first time, attempt to quantify the spatially inhomogeneous reionization from the wide-field survey with Subaru HSC.

\begin{enumerate}
    \item We simultaneously detect 189 LAEs and 179 LBGs at $z \sim 6.6$ in the COSMOS field with large FoV ($\sim1.5\,\mathrm{deg}^2$) HSC
    observations. The newly installed filter IB945 makes it possible to detect LBGs at similar redshift to LAEs. The surface number densities of our LAE and LBG samples are consistent with the previous studies.
    \item Based on the state-of-the-art simulation of reionization, the observed $n(\mathrm{LAE})/n(\mathrm{LBG})$, $0.84 ^{+0.23}_{-0.27}$, puts a constraint on the average neutral fraction in the FoV as $x_\mathrm{HI}<0.4$, which is consistent with previous studies.
    \item By comparing the density distribution of the LAEs and LBGs, we detect a spatial variation over a factor of three in $n(\mathrm{LAE})/n(\mathrm{LBG})$.
    Our model predicts that the spatial variation of $n(\mathrm{LAE})/n(\mathrm{LBG})$ corresponds to the spatial variation of the neutral fraction. i.e., the patchy reionization topology. This implies reionization is proceeding in the high LAE density region in our observation, while it is delayed in the low LAE density region.
    \item Based on the model, we conclude that the observed large scatter of $n(\mathrm{LAE})/n(\mathrm{LBG})$ can either be explained by the reionization topology or the intrinsic large-scale structure, and it may be difficult to distinguish them with the current survey depth.
    \item LAEs in the high LAE density region{s} are found to be more populate high $\mathrm{EW}_0$. The result supports that the observed $n(\mathrm{LAE})/n(\mathrm{LBG})$ is more or less driven by the neutral fraction, though the statistical significance is not high.
\end{enumerate}

\begin{acknowledgments}
    T. Y. is supported by International Graduate Program for Excellence in Earth-Space Science (IGPEES). N. K. is supported by JSPS grant 21H04490.
    {A. K. I. is supported by JSPS grant 21H04489.}
    {K. I. acknowledges support from JSPS grant 20J12461.}
    {R. M. acknowledges a Japan Society for the Promotion of Science (JSPS) Fellowship at Japan and the JSPS KAKENHI grant No. JP18J40088.}

    This paper is based on data collected at the Subaru Telescope and retrieved from the HSC data archive system, which is operated by Subaru Telescope and Astronomy Data Center (ADC) at NAOJ. Data analysis was in part carried out with the cooperation of Center for Computational Astrophysics (CfCA) at NAOJ.  We are honored and grateful for the opportunity of observing the Universe from Maunakea, which has the cultural, historical and natural significance in Hawaii.

    The Hyper Suprime-Cam (HSC) collaboration includes the astronomical communities of Japan and Taiwan, and Princeton University.  The HSC instrumentation and software were developed by the National Astronomical Observatory of Japan (NAOJ), the Kavli Institute for the Physics and Mathematics of the Universe (Kavli IPMU), the University of Tokyo, the High Energy Accelerator Research Organization (KEK), the Academia Sinica Institute for Astronomy and Astrophysics in Taiwan (ASIAA), and Princeton University.  Funding was contributed by the FIRST program from the Japanese Cabinet Office, the Ministry of Education, Culture, Sports, Science and Technology (MEXT), the Japan Society for the Promotion of Science (JSPS), Japan Science and Technology Agency  (JST), the Toray Science  Foundation, NAOJ, Kavli IPMU, KEK, ASIAA, and Princeton University.

    This paper makes use of software developed for Vera C. Rubin Observatory. We thank the Rubin Observatory for making their code available as free software at http://pipelines.lsst.io/.

    The Pan-STARRS1 Surveys (PS1) and the PS1 public science archive have been made possible through contributions by the Institute for Astronomy, the University of Hawaii, the Pan-STARRS Project Office, the Max Planck Society and its participating institutes, the Max Planck Institute for Astronomy, Heidelberg, and the Max Planck Institute for Extraterrestrial Physics, Garching, The Johns Hopkins University, Durham University, the University of Edinburgh, the Queen’s University Belfast, the Harvard-Smithsonian Center for Astrophysics, the Las Cumbres Observatory Global Telescope Network Incorporated, the National Central University of Taiwan, the Space Telescope Science Institute, the National Aeronautics and Space Administration under grant No. NNX08AR22G issued through the Planetary Science Division of the NASA Science Mission Directorate, the National Science Foundation grant No. AST-1238877, the University of Maryland, Eotvos Lorand University (ELTE), the Los Alamos National Laboratory, and the Gordon and Betty Moore Foundation.

    This research has made use of NASA’s Astrophysics Data System.

\end{acknowledgments}

\software{
    adstex (\url{https://github.com/yymao/adstex}),
    Astropy \citep{astropy13, astropy18},
    Datashader (\url{https://datashader.org/}),
    Jupyter \citep{jupyter},
    Matplotlib \citep{matplotlib},
    Numpy \citep{numpy},
    pandas \citep{pandas, McKinney10},
    SciPy \citep{scipy},
    uncertainties (\url{http://pythonhosted.org/uncertainties/})
}

\bibliography{chorus_iv_v4}{}
\bibliographystyle{aasjournal}

\appendix
\section{Robustness of the relation between $n(\mathrm{LAE})/n(\mathrm{LBG})$ and $x_\mathrm{HI}$}\label{append:nratio_xhi_relation}
\begin{figure}
    \centering
    \includegraphics[width=3.2in]{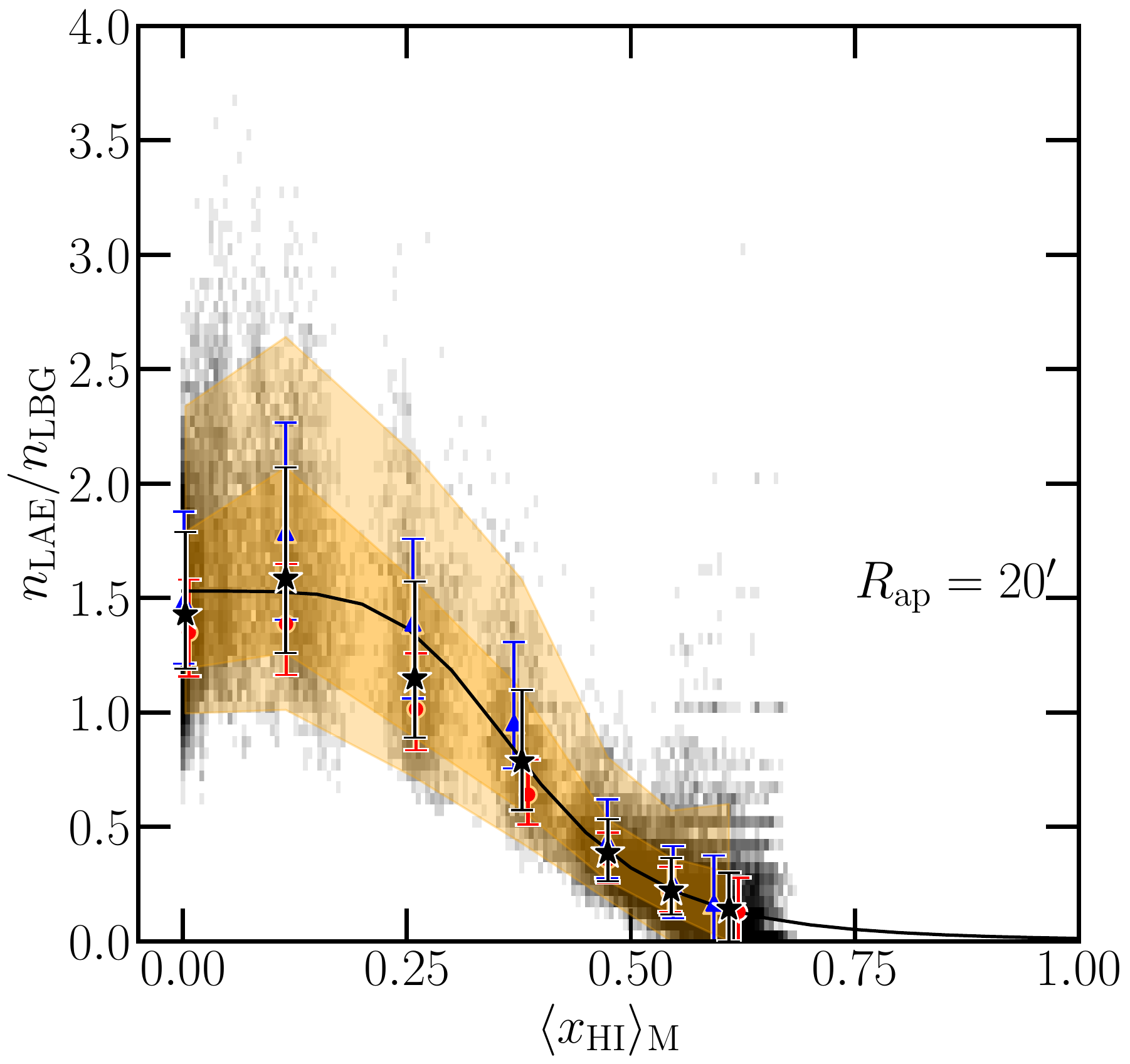}
    \caption{The relation between $x_\mathrm{HI}$ and $n(\mathrm{LAE})/n(\mathrm{LBG})$ in the simulation data with $\Delta z = 0.1$ top-hat selection function for LAEs and $\Delta z = 0.3$ for LBGs. Colors are the same as Figure \ref{fig:nratio_xHI_relation}.}\label{fig:nratio_xhi_relation_tophat}
\end{figure}

In Section \ref{sec:neutralfraction}, we derive the relation between the number ratio $n(\mathrm{LAE})/n(\mathrm{LBG})$ and the neutral fraction $x_\mathrm{HI}$ using the observed selection functions described in Section \ref{sec:completeness}.
To see the robustness of the trend, we here make a simpler assumption on the selection function.
we use a $\Delta z = 0.1$ top-hat selection function for LAEs and $\Delta z=0.3$ for LBGs.
Figure \ref{fig:nratio_xhi_relation_tophat} shows the correlation between $x_\mathrm{HI}$ and $n(\mathrm{LAE})/n(\mathrm{LBG})$ with this selection.
When comparing $n(\mathrm{LAE})/n(\mathrm{LBG})$ between the simulation and observation, we need to correct the difference in selection methods.
In the simulation data, we select LAEs and LBGs with the top-hat selection and the photometric selection in the same as the observational data.
We use a ratio of the number of the objects with the top-hat selection to that with photometric selection as a correction factor between the simulation and the observation.
We derive observed $n(\mathrm{LAE})/n(\mathrm{LBG})$ with the top-hat selection by multiplying the correction factor to the observed number of the objects and derive the neutral fraction in the same way as Section \ref{sec:neutralfraction}.
The neutral fraction is $x_\mathrm{HI} < 0.4$, which is consistent with the result of Section \ref{sec:neutralfraction}.
We confirm that the effect of the difference in selection methods is negligible and that our results are robust.

\end{document}